\documentclass[preprint,10pt]{elsarticle}
\usepackage{layout}
\usepackage{amsmath,amsfonts,amssymb}
\usepackage{xcolor}
\usepackage{hyperref}
\usepackage{epsfig}
\usepackage{todonotes}
\usepackage{algorithm}
\usepackage{algpseudocode}
\usepackage{lscape}
\usepackage{subcaption}
\usepackage{multirow}

\usepackage{pgfplots}
\usepackage[font=sl,labelfont=normal,size=small]{caption}
\usepackage{rotating}
\definecolor{lightgray}{gray}{0.95}

\usepackage[a4paper,text={6.0in,9.0in},centering,includefoot,foot=0.6in]{geometry} 

\date{\today}

\title{The communication-hiding pipelined BiCGStab method for the parallel solution of large unsymmetric linear systems}

\author[ua]{\underline{S. Cools}} \ead{siegfried.cools@uantwerpen.be}
\author[ua]{W. Vanroose} \ead{wim.vanroose@uantwerpen.be}
\address[ua]{Applied Mathematics Research Group, Department of Mathematics and Computer Science, University of Antwerp, Middelheimlaan 1, B-2020 Antwerp, Belgium. Contact: \underline{siegfried.cools@uantwerp.be}, wim.vanroose@uantwerp.be\vspace{-0.6cm}} 

\begin{document}

\begin{abstract} 
	A High Performance Computing alternative to traditional Krylov subspace methods, pipelined Krylov subspace solvers offer 
	better scalability in the strong scaling limit compared to standard Krylov subspace methods for large and sparse linear 
	systems. The typical synchronization bottleneck is mitigated by overlapping time-consuming global communication phases with 
	local computations in the algorithm. This paper describes a general framework for deriving the pipelined 
	variant of any Krylov subspace algorithm. The proposed framework was  
	implicitly used to derive the pipelined Conjugate Gradient (p-CG) method in \emph{Hiding global synchronization latency 
	in the preconditioned Conjugate Gradient algorithm} by P.~Ghysels and W.~Vanroose, Parallel Computing, 40(7):224--238, 2014.
	The pipelining framework is subsequently illustrated by formulating a pipelined version of the BiCGStab method 
	for the solution of large unsymmetric linear systems on parallel hardware. A residual 
	replacement strategy is proposed to account for the possible loss of attainable accuracy and robustness 
	by the pipelined BiCGStab method. It is shown that the pipelined algorithm improves scalability 
	on distributed memory machines, leading to significant speedups compared to standard preconditioned BiCGStab.
\end{abstract}


\maketitle

\section{Introduction} \label{sec:introduction} 

At present, Krylov subspace methods fulfill the role of standard linear algebra solution methods in many high-performance computing (HPC) applications where large and sparse linear systems need to be solved. The Conjugate Gradient (CG) method \cite{hestenes1952methods} can be considered as the earliest member of this well-known class of iterative solvers. However, the CG method is restricted to the solution of symmetric and positive definite (SPD) systems. Several variants of the CG method have been proposed that allow for the solution of more general classes of unsymmetric and indefinite linear systems. These include e.g.~the Bi-Conjugate Gradient (BiCG) method \cite{fletcher1976conjugate}, the Conjugate Gradient Squared (CGS) method \cite{sonneveld1989cgs}, and the widely used BiCGStab method which was introduced as a smoother converging version of the  aforementioned methods by H.A.~Van der Vorst in 1992 \cite{van1992bi}.

Motivated by the vast numbers of processors in contemporary petascale and future exascale HPC hardware, research on the scalability of Krylov subspace methods on massively parallel hardware has become increasingly prominent over the last years. The practical importance of Krylov subspace methods for solving sparse linear systems is reflected in the High Performance Conjugate Gradient (HPCG) benchmark used for ranking HPC systems introduced in 2013 \cite{dongarra2013toward, dongarra2015hpcg}. This new ranking is based on sparse matrix-vector computations and data access patterns, rather than the dense matrix algebra used in the traditional High Performance LINPACK (HPL) benchmark. 

Krylov subspace algorithms are build from three basic components, namely: \emph{sparse matrix-vector products} (\textsc{spmv}s) $Ax$, \emph{dot-products} (\textsc{dot-prod}s) or inner products $(x,y)$ between to vectors, and \textsc{axpy} operations $x \leftarrow a x + y$ that combine \emph{scalar multiplication} $ax$ and \emph{vector addition} $x+y$.
Single-node Krylov solver performance is dominated by the computation of the \textsc{spmv}, which has the highest floating-point operation (\textsc{flop}) count of the algorithmic components. However, the main bottleneck for efficient parallelization is typically not the easy-to-parallelize \textsc{spmv} application, but the dot-product computation that requires a global reduction over all processors. These global communications are commonly performed through a 2-by-2 reduction tree, which requires $\mathcal{O}\left(\log_2(P)\right)$ time, where $P$ is the number of processors in the machine, see \cite{ghysels2013hiding}. The resulting global synchronization phases cause severe communication overhead on parallel machines. Hence, dot-product computations are communication-bound, and are thus effectively the most time-consuming component of Krylov subspace algorithms for large scale linear systems.

A significant amount of research has been devoted to the reduction and elimination of the synchronization bottlenecks in Krylov subspace methods over the last decades. The idea of reducing the number of global communication phases and hiding the communication latency in Krylov subspace methods on parallel computer architectures was first presented by Chronopoulos and Gear in \cite{chronopoulos1989s} and further elaborated in a variety of papers, among which \cite{barrett1994templates,de1991parallel,erhel1995parallel}. In 2002, Yang et al.~\cite{yang2002improved2,yang2002improved,yang2003improved} proposed the so-called \emph{improved} versions of the standard BiCG, BiCGStab and CGS algorithms, which reduce the number of global reductions to only one per iteration. Furthermore, the $s$-step formulation of Krylov subspace methods \cite{carson2015communication,carson2014residual,carson2013avoiding,chronopoulos1991s,chronopoulos2010block,chronopoulos1996parallel} allows to reduce the total number of global synchronization points by a factor of $s$.

In addition to avoiding communication by reducing the number of global synchronization points, research on hiding global communication latency by overlapping communication and computations was performed by, among others, Demmel et al.~\cite{demmel1993parallel}, De Sturler et al.~\cite{de1995reducing} and Ghysels et al.~\cite{ghysels2013hiding,ghysels2014hiding}. The latter introduces the class of pipelined Krylov subspace methods, which in addition to \emph{removing} costly global synchronization points (communication-avoiding), aim at \emph{overlapping} the remaining global communication phases by the \textsc{spmv} and the preconditioner application (communication-hiding). In this way idle core time is minimized by performing useful computations simultaneously to the time-consuming global communication phases, cf.~\cite{eller2015non,sanan2016pipelined}.

The current paper aims to extend the work on pipelined Krylov subspace methods. The article is structured as follows. Section \ref{sec:framework} introduces a high-level framework for the derivation of a pipelined method starting from the original Krylov subspace algorithm. Section \ref{sec:pipe-bicgstab} illustrates the use of this framework by deriving the new pipelined BiCGStab (p-BiCGStab) method. The inclusion of a preconditioner, which is conveniently simple for pipelined Krylov subspace methods, is also discussed here. Experimental results with the p-BiCGStab method on a wide range of problems are proposed in Section \ref{sec:numerical} to validate the mathematical equivalence between the traditional and pipelined methods. A residual replacement strategy is proposed to improve the numerical accuracy and robustness of the pipelined BiCGStab solver. Section \ref{sec:parallel} compares the parallel performance of the traditional and pipelined BiCGStab methods on a moderately-sized cluster and comments on the numerical accuracy of the pipelined BiCGStab solution. Finally, conclusions are formulated in Section \ref{sec:conclusions}.

\section{General framework for deriving a pipelined Krylov subspace algorithm} \label{sec:framework}

This section provides the outline of a generic work plan for deriving the pipelined version of any traditional Krylov subspace method. The framework is based on the two main properties of the pipelining technique: \emph{avoiding communication}, achieved by reducing the number of global reductions where possible, and \emph{hiding communication} by overlapping local computational work (\textsc{axpy}s, \textsc{spmv}s) with the global communication required for the composition of dot-products. The framework below was implicitly used for the derivation of existing pipelined algorithms, such as the pipelined p-CG and p-CR methods in \cite{ghysels2014hiding}, as well as the $l^1$-GMRES algorithm proposed in \cite{ghysels2013hiding}. 
The different steps in the theoretical framework are illustrated by its application to the pipelined CG method.

\begin{algorithm}[t!]
  \caption{Standard Krylov subspace method \hfill \textbf{Step 1: Avoiding communication}}
  \label{algo::krylov1}
  \begin{algorithmic}[1]
  	\Function{krylov}{$A$, $b$, $x_0$}
    \State \ldots  				
					\hfill \fcolorbox{black}{lightgray}{\makebox[0.8cm]{\textbf{1(a)}}  \makebox[2.2cm]{identification}  	\quad \makebox[1.0cm]{~~~~}}
    \For{$i = 0,1,2, \dots$}
    \State \ldots 																																				
					\hfill \fcolorbox{black}{lightgray}{\makebox[0.8cm]{\textbf{1(d)}} \makebox[2.2cm]{new \textsc{spmv}s}} 	\quad \makebox[1.0cm]{(S2)}
		\State \textbf{begin reduction} \textsc{dot-prod}s $\left(  x ,  y \right)$ \textbf{end reduction}		
					\hfill 	\quad \makebox[1.0cm]{(R1)}
		\State \ldots
		\State \textbf{computation} \textsc{spmv}s $ A x  $ 						
					\hfill \fcolorbox{black}{lightgray}{\makebox[0.8cm]{\textbf{1(b)}} \makebox[2.2cm]{recurrences}} 					\quad \makebox[1.0cm]{(S1)}
		\State \ldots 
		\State \textbf{begin reduction} \textsc{dot-prod}s $\left( x , y \right)$ 	\textbf{end reduction}	
					\hfill \fcolorbox{black}{lightgray}{\makebox[0.8cm]{\textbf{1(c)}} \makebox[2.2cm]{reformulation}} 					\quad \makebox[1.0cm]{(R2)}
		\State \ldots
    \EndFor
    \EndFunction
  \end{algorithmic}
\end{algorithm}

\begin{algorithm}[t]
  \caption{Standard CG (preconditioned)}
  \label{algo::pcg}
  \begin{algorithmic}[1]
  	\Function{prec-cg}{$A$, $M^{-1}$, $b$, $x_0$}
    \State $r_0 := b - Ax_0$; $u_0:= M^{-1} r_0$; $p_0 := u_0$  \label{al:residual}
    \For{$i = 0,1,2, \dots$}
    \State \textbf{computation} $s_i := Ap_{i}$ \label{al:matvec}
		\State \textbf{begin reduction} $\left( s_i, p_i \right)$ \textbf{end reduction}
    \State $\alpha_{i} := \left( r_i, u_i \right) / \left( s_i, p_i \right)$ \label{al:alpha}
    \State $x_{i+1} := x_i + \alpha_{i} p_i$
    \State $r_{i+1} := r_i - \alpha_{i} s_i$
    \State \textbf{computation} $u_{i+1} := M^{-1} r_{i+1}$
		\State \textbf{begin reduction} $\left( r_{i+1}, u_{i+1} \right)$ \textbf{end reduction}
    \State $\beta_{i+1} := \left( r_{i+1}, u_{i+1} \right) / \left( r_i, u_i \right)$
    \State $p_{i+1} := u_{i+1} + \beta_{i+1} p_i$ 
    \EndFor
    \EndFunction
  \end{algorithmic}
\end{algorithm}

\subsection{Step 1: Avoiding communication}

The primary goal in the first step is the merging of global reduction phases (\textsc{dot-prod}s) to reduce the number of global synchronization points, and hence, the overall time spent in communication. Algorithm \ref{algo::krylov1} shows a schematic illustration of the original Krylov subspace algorithm (left) and indicates the operations that are performed in Step 1 (right) to obtain the communication-avoiding (or CA, for short) Krylov subspace method. The latter is presented schematically in Algorithm \ref{algo::krylov2} (left). The following sequence of steps (a) to (d) may be repeated any number of times to merge multiple global reduction phases, if required. 

	\begin{itemize} \setlength\itemsep{0.0em}
	\item[1(a)] Identify two global reduction phases that are candidates for merging. 
	The global reduction phase that is located before the intermediate \textsc{spmv} computation (S1) is denoted (R1) and the subsequent reduction phase that is executed after the \textsc{spmv} is called (R2). 
	\\ \vspace{-0.3cm} \\
	{\underline{E.g.:} CG: reduction (R1) is found on line 10 in Algorithm \ref{algo::pcg}, 
	while (R2) is on line 5. The intermediate \textsc{spmv} is located on line 4.}
	
	\item[1(b)] Introduce a recurrence for the \textsc{spmv} computation (S1) that is computed 
	in between the global reduction phases (R1) and (R2). To achieve this, substitute the \textsc{spmv} vector by its 
	respective recurrence, which is computed before (S1). Define new \textsc{spmv} variables 
	(S2) where needed and compute these \textsc{spmv}s directly below their corresponding 
	recursive vector definitions.
	\\ \vspace{-0.3cm} \\
	{\underline{E.g.:} CG: the \textsc{spmv} (S1) on line 4 of Alg.~\ref{algo::pcg} 
	is replaced by $s_i = A u_i + \beta_i s_{i-1}$. 	The new \textsc{spmv} variable $w_i = A u_i$ 
	(S2) is introduced directly below the definition of $u_{i+1}$ on line 9.}
	
	\item[1(c)] Use the new recurrence defined in (b) to reformulate the 
	dot-products of the (R2) reduction phase independently of any intermediate variables that 
	are computed between (R1) and (R2).
	\\ \vspace{-0.3cm} \\
	{\underline{E.g.:} CG: occurrences of $s_i$ and $p_i$ in (R2) on line 5 of 
	Alg.~\ref{algo::pcg} are replaced by their respective recurrences (see lines 4 and 12). The expression for $\alpha_i$ 
	is reformulated accordingly.}
	
	\item[1(d)] Move the (R2) global reduction phase upward to merge it with the (R1) global 
	reduction into a single global communication phase.
	\\ \vspace{-0.3cm} \\
	{\underline{E.g.:} CG: the reduction (R2) on line 5 is independent of the intermediate 
	variables $s_i$ and $p_{i+1}$. It can thus be moved upward (to the previous iteration)	to compute 
	$\alpha_{i+1}$ right below $\beta_{i+1}$.}
	
	\end{itemize}
	
\noindent {\underline{E.g.:} CG-CG: Step 1 results in the communication avoiding Conjugate Gradient method by 
Chronopoulos and Gear \cite{chronopoulos1989s}, denoted as CG-CG for short, outlined in Algorithm \ref{algo::pcgcg}.}\\ 

For preconditioned Krylov subspace methods the \textsc{spmv} phases (S1) and (S2) may 
also include the (right) preconditioner application, which is typically computed right before the 
\textsc{spmv}. 
This is exemplified by the CG algorithms, and is illustrated for the BiCGStab method in Section \ref{sec:precpipebicgstab}.

\begin{algorithm}[t!]
  \caption{CA-Krylov subspace method \hfill \textbf{Step 2: Hiding communication}}
  \label{algo::krylov2}
  \begin{algorithmic}[1]
  	\Function{ca-krylov}{$A$, $b$, $x_0$}
    \State \ldots
					\hfill \fcolorbox{black}{lightgray}{\makebox[0.8cm]{\textbf{2(a)}}  \makebox[2.2cm]{identification}  	\quad \makebox[1.0cm]{~~~~}}
    \For{$i = 0,1,2, \dots$}
		\State \ldots
		\State \textbf{computation} \textsc{spmv}s $A x$ 						
					\hfill \fcolorbox{black}{lightgray}{\makebox[0.8cm]{\textbf{2(b)}} \makebox[2.2cm]{recurrences}} 					\quad \makebox[1.0cm]{(S2)}
		\State \ldots 
		\State \textbf{begin reduction} \textsc{dot-prod}s $\left( x, y \right)$ \textbf{end reduction}
					\hfill \fcolorbox{black}{lightgray}{\makebox[0.8cm]{\textbf{2(c)}} \makebox[2.2cm]{reformulation}} 					\quad \makebox[1.0cm]{(R1)}
		\State \ldots
		\State \ldots																																					
					\hfill \fcolorbox{black}{lightgray}{\makebox[0.8cm]{\textbf{2(d)}} \makebox[2.2cm]{new \textsc{spmv}s}}	\quad \makebox[1.0cm]{(S3)}
    \EndFor
    \EndFunction
  \end{algorithmic}
\end{algorithm}

\begin{algorithm}[t]
  \caption{Chronopoulos \& Gear CG (preconditioned)}
  \label{algo::pcgcg}
    \begin{algorithmic}[1]
    	\Function{prec-cg-cg}{$A$, $M^{-1}$, $b$, $x_0$}
      \State $r_0 := b - Ax_0$; $u_0 := M^{-1}r_0$; $w_0 := Au_0$
      \State $\alpha_0 := (r_0,u_0)/(w_0,u_0)$; $\beta_0:=0$; $\gamma_0 := (r_0,u_0)$
      \For{$i = 0,1,2, \dots$}
      \State $p_i := u_i + \beta_i p_{i-1}$
      \State {$s_i := w_i + \beta_i s_{i-1}$}
      \State {$x_{i+1} := x_i + \alpha_i p_i$}
      \State {$r_{i+1} := r_i - \alpha_i s_i$}
      \State {\textbf{computation} $u_{i+1} := M^{-1} r_{i+1}$}
      \State {\textbf{computation} $w_{i+1} := Au_{i+1}$}
      \State {\textbf{begin reduction} $\gamma_{i+1} := (r_{i+1},u_{i+1})$; $\delta := (w_{i+1},u_{i+1})$  \textbf{end reduction}}
      \State {$\beta_{i+1} := \gamma_{i+1}/\gamma_i$}
      \State {$\alpha_{i+1} := (\delta/\gamma_{i+1}-\beta_{i+1}/\alpha_i)^{-1}$}
      \EndFor
			\EndFunction
    \end{algorithmic}
\end{algorithm}

\subsection{Step 2: Hiding communication}

The aim of the second step is the simultaneous execution (overlapping) of the global reduction communication phases with independent local \textsc{spmv} computations to hide communication time behind useful computations and hence minimize worker idling. Step 2 is illustrated in Algorithm \ref{algo::krylov2} and exemplified by the CG-CG method in Algorithm \ref{algo::pcgcg}. Starting from the communication-avoiding Krylov subspace method that is represented schematically by Alg.~\ref{algo::krylov2} (left), the operations (a)-(d) in Step 2 required to obtain the pipelined Krylov subspace method are indicated in Alg.~\ref{algo::krylov2} (right). The following reformulations ultimately leads to Alg.~\ref{algo::krylov3}-\ref{algo::ppipe-cg}.

\begin{algorithm}[t!]
  \caption{Pipelined Krylov subspace method \hfill \textbf{Final pipelined algorithm}}
  \label{algo::krylov3}
  \begin{algorithmic}[1]
  	\Function{pipe-krylov}{$A$, $b$, $x_0$}
    \State \ldots
    \For{$i = 0,1,2, \dots$}
		\State \ldots				
		\State \textbf{begin reduction} \textsc{dot-prod}s $\left( x , y \right)$ 		\hfill \quad \makebox[1.0cm]{(R1)}
		\State \textbf{computation} \textsc{spmv}s $A x$ 						
		\hfill \fcolorbox{black}{lightgray}{\makebox[2.2cm]{overlapping}} 					\quad \makebox[1.0cm]{(S3)}
		\State \textbf{end reduction}
		\State \ldots																																					
    \EndFor
    \EndFunction
  \end{algorithmic}
\end{algorithm}

\begin{algorithm}[t]
  \caption{Pipelined CG (preconditioned)}
  \label{algo::ppipe-cg}
  \begin{algorithmic}[1]
  	\Function{prec-p-cg}{$A$, $M^{-1}$, $b$, $x_0$}
    \State $r_0 := b - Ax_0$; $u_0:= M^{-1} r_0$; $w_0 := Au_0$
    \For{$i = 0,1,2,\dots$}
    \State \textbf{begin reduction} $\gamma_i :=(r_i,u_i)$; $\delta := (w_i,u_i)$
    \State \textbf{computation} $m_i := M^{-1} w_i$
    \State \textbf{computation} $n_i := A m_i$
		\State \textbf{end reduction}
    \If{$i>0$}
    \State $\beta_i := \gamma_i/\gamma_{i-1}$; $\alpha_i := (\delta/\gamma_i - \beta_i/\alpha_{i-1})^{-1}$
    \Else
    \State $\beta_i :=0$; $\alpha_i := \gamma_i/\delta$
    \EndIf
    \State $z_i := n_i + \beta_i z_{i-1}$
    \State $q_i := m_i + \beta_i q_{i-1}$
    \State $s_i := w_i + \beta_i s_{i-1}$
    \State $p_i := u_i + \beta_i p_{i-1}$
    \State $x_{i+1} := x_i + \alpha_i p_i$
    \State $r_{i+1} := r_i - \alpha_i s_i$
    \State $u_{i+1} := u_i - \alpha_i q_i$
    \State $w_{i+1} := w_i - \alpha_i z_i$
    \EndFor
    \EndFunction
  \end{algorithmic}
\end{algorithm}

	\begin{itemize} \setlength\itemsep{0.0em}
	\item[2(a)] Following Step 1, each \textsc{spmv} phase (S2) is followed by a corresponding 
	global reduction phase (R1). These \textsc{spmv}/\textsc{dot-prod} pairs are possibly separated 
	by intermediate \textsc{axpy} operations. 
	\\ \vspace{-0.3cm} \\
	{\underline{E.g.:} CG-CG: in Algorithm \ref{algo::pcgcg} the \textsc{spmv} (S2) on line 9-10 is 
	directly followed by the global reduction phase (R1) required for the \textsc{dot-prod}s on line 11.}
	
	\item[2(b)] Introduce a recurrence for the \textsc{spmv} vector in (S2) through substitution 
	by its recursive characterization, which is computed before (S2). Define new \textsc{spmv} 
	variables where needed (S3).
	\\ \vspace{-0.3cm} \\
	{\underline{E.g.:} CG-CG: the \textsc{spmv} (S2) on line 10 in Alg.~\ref{algo::pcgcg} 
	is replaced by the recurrence $w_{i+1} = w_i -\alpha_i A M^{-1} s_i$. Here the \textsc{spmv} variable is 
	defined as $z_i = A M^{-1} s_i$. However, using the recurrence for $s_i$, $z_i$ can in turn be computed recursively as 
	$z_i = A M^{-1} w_i + \beta_i z_{i-1}$. The new \textsc{spmv} $n_i = A M^{-1} w_i$ (S3) is introduced 
	below the recursive definition of $w_{i+1}$ on line 10.}
		
	\item[2(c)] Use the new recursive definition obtained in (b) to reformulate the dot-products of 
	the (R1) reduction phase independently of the new \textsc{spmv} variables that are computed in (S3).
	\\ \vspace{-0.3cm} \\
	{\underline{E.g.:} CG-CG: the (R1) phase on line 11 in Alg.~\ref{algo::pcgcg} is independent of the \textsc{spmv} variable $n_i$.}
	
	\item[2(d)] Insert the new \textsc{spmv} variables (S3) defined in (b) right after the global reduction 
	phase (R1). These \textsc{spmv} computations can now be overlapped with the global synchronization (R1), 
	since they do not make use of the results computed in the (R1) phase.
	\\ \vspace{-0.3cm} \\
	{\underline{E.g.:} CG-CG: the \textsc{spmv} $n_i = A M^{-1}w_i$ (S3) is inserted below the (R1) global 
	reduction phase on line 11. Since there are no dependencies between these phases, they can be overlapped.}
	
	\end{itemize}
	
The resulting pipelined algorithm is shown in Algorithm \ref{algo::krylov3}. The proposed framework allows for the derivation of a pipelined Krylov subspace method with a length-one pipeline, similar to the p-CG and p-CR methods described in \cite{ghysels2014hiding}. A general framework for derivation of methods with longer pipeline lengths, cf.~the p-GMRES($l$) method described in \cite{ghysels2013hiding}, is highly non-trivial and is left as future work.
\\ \vspace{-0.3cm} \\
{\underline{E.g.:} p-CG: Applying Step 2 to the CG-CG algorithm yields the communication-hiding pipelined CG method (p-CG) by 
Ghysels et al.~\cite{ghysels2014hiding}. This method is shown in Algorithm \ref{algo::ppipe-cg}.}

\section{Derivation of the pipelined BiCGStab algorithm} \label{sec:pipe-bicgstab}

\subsection{The standard BiCGStab algorithm}

The traditional Biconjugate Gradient Stabilized method (BiCGStab), shown in Algorithm \ref{algo::bicgstab1}, was developed as a fast and smoothly converging variant of the BiCG and CGS methods \cite{van1992bi}. It presently serves as the standard \emph{workhorse} Krylov subspace method for the iterative solution of nonsymmetric linear systems $A x = b$, where $A$ is a general real or complex matrix with generic spectral properties, be it positive definite, positive or negative semidefinite, or indefinite.

Alg.~\ref{algo::bicgstab1} performs two \textsc{spmv} applications (lines 4 and 8) and a total of 4 recursive 
vector updates (lines 7, 11, 12 and 15) in each iteration. When implemented on a parallel machine, the \textsc{axpy}s used to compute the recurrences are local operations, requiring no communication between individual workers.
The \textsc{spmv}s can be considered semi-local operations, since only limited communication between neighboring workers is required for boundary elements.
In the case of stencil application, or the application of the banded sparse matrix structures that typically result from the discretization of PDEs, 
data locality can be exploited to ensure limited communication is required for computing the \textsc{spmv}. Hence, the \textsc{spmv}
operations are considered to be primarily compute-bound.
The traditional BiCGStab algorithm additionally features three global reduction steps to compute the dot-products required in the calculation of the 
scalar variables $\alpha_i$ (line 5-6), $\omega_i$ (line 9-10) and $\beta_i$ (line 13-14). On parallel machines these dot-products require global 
communication among all workers to assemble the locally computed dot-product fractions and redistribute the final scalar result to all workers. 

\begin{algorithm}[t]
  \caption{Standard BiCGStab}
  \label{algo::bicgstab1}
  \begin{algorithmic}[1]
  	\Function{bicgstab}{$A$, $b$, $x_0$}
    \State $r_0 := b - Ax_0$; $p_0 := r_0$  
    \For{$i = 0,1,2, \dots$}
    \State \textbf{computation} $s_i := Ap_{i}$ 
		\State \textbf{begin reduction} $\left( r_0, s_i \right)$ \textbf{end reduction}
    \State $\alpha_i := \left( r_0, r_i \right) / \left( r_0, s_i \right)$ 
		\State $q_i := r_i - \alpha_i s_i$
		\State \textbf{computation} $y_i := A q_i$
		\State \textbf{begin reduction} $\left( q_i, y_i \right)$; $\left( y_i, y_i \right)$ \textbf{end reduction}
		\State $\omega_i := \left( q_i, y_i \right)/\left( y_i, y_i \right)$
    \State $x_{i+1} := x_i + \alpha_{i} p_i + \omega_i q_i$
    \State $r_{i+1} := q_i - \omega_{i} y_i$
		\State \textbf{begin reduction} $\left( r_0, r_{i+1} \right)$ \textbf{end reduction}
    \State $\beta_i := \left( \alpha_i / \omega_i \right)  \left( r_0, r_{i+1} \right) / \left( r_0, r_i \right)$
    \State $p_{i+1} := r_{i+1} + \beta_i \left( p_i - \omega_i s_i \right)$ 
    \EndFor
    \EndFunction
  \end{algorithmic}
\end{algorithm}


\subsection{Step 1: Avoiding global communication: towards CA-BiCGStab} \label{sec:avoiding}

Starting from the original BiCGStab Algorithm \ref{algo::bicgstab1}, first the number of global communication phases is reduced. 
To this aim the dot-product for the computation of $\alpha_i$ (line 5) is merged with
the global reduction phase required to compute $\beta_i$ (line 13). Rewriting the intermediate \textsc{spmv} $s_i = A p_i$ by using the recurrence for $p_i$ on line 15, we obtain:
\begin{align} \label{eq:s_i}
	s_i = A p_i 	&= A \left(r_i + \beta_{i-1} \left(p_{i-1} - \omega_{i-1} s_{i-1} \right)\right) \notag \\
								&= w_i + \beta_{i-1} \left( s_{i-1} - \omega_{i-1} z_{i-1} \right),
\end{align}
where we use that $s_{i-1} = Ap_{i-1}$, and the auxiliary variables $w_i = Ar_i$ and $z_i = As_i$ are defined.
Subsequently, the new recurrence for $s_i$ \eqref{eq:s_i} is applied to rewrite the dot-product $(r_0,s_i)$ (line 5), 
so that $\alpha_i$ becomes independent of the intermediate variables $s_i$ and $p_{i+1}$, i.e.,
\begin{align} \label{eq:r_0_s_i}
	(r_0,s_i) &= (r_0,w_i + \beta_{i-1} \left( s_{i-1} - \omega_{i-1} z_{i-1} \right)) \notag \\
						&= (r_0,w_i) + \beta_{i-1} (r_0, s_{i-1}) -\beta_{i-1} \omega_{i-1} (r_0,z_{i-1}).
\end{align}
By substituting \eqref{eq:r_0_s_i} into the definition of $\alpha_i = \left( r_0, r_i \right) / \left( r_0, s_i \right)$, the new expression for $\alpha_i$ becomes
\begin{equation} \label{eq:alpha_i}
	\alpha_i = \frac{\displaystyle(r_0,r_i)}{\displaystyle(r_0,w_i) + \beta_{i-1} (r_0, s_{i-1}) -\beta_{i-1} \omega_{i-1} (r_0,z_{i-1})},
\end{equation}
which requires the dot-products $(r_0,r_i)$, $(r_0,w_i)$, $(r_0, s_{i-1})$ and $(r_0,z_{i-1})$, but no longer uses the dot-product $(r_0,s_i)$.
Since $\alpha_i$ is now independent of the intermediate variables $s_i$ and $p_{i+1}$, it can be moved upward, and its global reduction phase can be merged with the global reduction required to compute $\beta_i$ (line 13).
The \textsc{spmv} $y_i = A q_i$ (line 8) can also be replaced by a recurrence by using the new variables $w_i = A r_i$ and $z_i = A s_i$, i.e.,
\begin{equation} \label{eq:y_i}
	y_i = Aq_i = A \left( r_i - \alpha_i s_i \right) = w_i - \alpha_i z_i.
\end{equation}
Hence, the total number of \textsc{spmv}s to be computed is two, similar to the original BiCGStab algorithm.
After a minor reordering of operations, this leads to the communication-avoiding version of BiCGStab shown in Algorithm \ref{algo::bicgstab2}. Note that this algorithm resembles the BiCGStab variant proposed in \cite{jacques1999electromagnetic}.

\begin{algorithm}[t]
  \caption{Communication-avoiding BiCGStab}
  \label{algo::bicgstab2}
  \begin{algorithmic}[1]
  	\Function{ca-bicgstab}{$A$, $b$, $x_0$}
    \State $r_0 := b - Ax_0$; $w_0 := A r_0$; $\alpha_0 := \left( r_0, r_0 \right) / \left( r_0, w_0 \right)$; $\beta_{-1} := 0$ 
    \For{$i = 0,1,2, \dots$}
		\State $p_i := r_i + \beta_{i-1} \left( p_{i-1} - \omega_{i-1} s_{i-1} \right)$ 
    \State $s_i := w_i + \beta_{i-1} \left( s_{i-1} - \omega_{i-1} z_{i-1} \right)$ 
		\State \textbf{computation} $z_i := A s_i$
		\State $q_i := r_i - \alpha_i s_i$
		\State $y_i := w_i - \alpha_i z_i$
		\State \textbf{begin reduction} $\left( q_i, y_i \right)$; $\left( y_i, y_i \right)$ \textbf{end reduction}
		\State $\omega_i := \left( q_i, y_i \right)/\left( y_i, y_i \right)$
    \State $x_{i+1} := x_i + \alpha_{i} p_i + \omega_i q_i$
    \State $r_{i+1} := q_i - \omega_{i} y_i$
		\State \textbf{computation} $w_{i+1} := A r_{i+1}$
		\State \textbf{begin reduction} $\left( r_0, r_{i+1} \right)$; $\left( r_0, w_{i+1} \right)$; $\left( r_0, s_i \right)$; $\left( r_0, z_i \right)$ \textbf{end reduction}
    \State $\beta_i := \left( \alpha_i / \omega_i \right)  \left( r_0, r_{i+1} \right) / \left( r_0, r_i \right)$
		\State $\alpha_{i+1} := \left( r_0, r_{i+1} \right) / \left( \left( r_0, w_{i+1} \right) + \beta_i \left( r_0, s_i \right) - \beta_i \omega_i \left( r_0, z_i \right) \right)$
    \EndFor
    \EndFunction
  \end{algorithmic}
\end{algorithm}

Although the total number of dot-products in Algorithm \ref{algo::bicgstab2} is two higher than in the original BiCGStab algorithm, the number of global reduction phases was reduced from three in Alg.~\ref{algo::bicgstab1} to only two in Alg.~\ref{algo::bicgstab2}. The first reduction phase in Alg.~\ref{algo::bicgstab2} (line 9) is unaltered compared to Alg.~\ref{algo::bicgstab1} (line 9). In the second global reduction the dot-products $\left( r_0, r_{i+1} \right)$, $\left( r_0, w_{i+1} \right)$, $\left( r_0, s_i \right)$ and $\left( r_0, z_i \right)$ are communicated simultaneously. This uses slightly more memory bandwidth, but leads to only one global synchronization point to compute both $\alpha_{i+1}$ and $\beta_i$ in Alg.~\ref{algo::bicgstab2} (line 14). 
The extra bandwidth usage for performing multiple dot-product communications in the same global synchronization is virtually negligible, since communication is limited to scalars only.

Note that the expression \eqref{eq:alpha_i} for $\alpha_{i+1}$ can alternatively be rewritten as 
\begin{equation} \label{eq:alpha_i_plus}
	\alpha_{i+1} = \left(\frac{\displaystyle 1}{\displaystyle \omega_i} + \frac{\displaystyle (r_0,w_{i+1})}{\displaystyle (r_0,r_{i+1})} - \beta_i \omega_i \frac{\displaystyle (r_0,z_i)}{\displaystyle (r_0, r_{i+1)}}\right)^{-1}.
\end{equation}
The expression for $\alpha_{i+1}$ above is equivalent to \eqref{eq:alpha_i} in exact arithmetic, but uses three dot-products instead of four. Since the number of global reductions remains unaffected by this operation, replacing the expression for $\alpha_{i+1}$ by \eqref{eq:alpha_i_plus} has no effect on the time spent in communication. Numerical experiments in finite precision arithmetic have shown that expression \eqref{eq:alpha_i} results in a more robust variant of the BiCGStab algorithm, in particular when combined with a residual replacement strategy, see Section \ref{sec:numerical}.

\subsection{Step 2: Hiding global communication: towards p-BiCGStab}

\begin{algorithm}[t]
  \caption{Pipelined BiCGStab}
  \label{algo::bicgstab3}
  \begin{algorithmic}[1]
  	\Function{p-bicgstab}{$A$, $b$, $x_0$}
    \State $r_0 := b - Ax_0$; $w_0 := A r_0$; $t_0 := A w_0$; $\alpha_0 := \left( r_0, r_0 \right) / \left( r_0, w_0 \right)$; $\beta_{-1} := 0$  
    \For{$i = 0,1,2, \dots$}
    \State $p_i := r_i + \beta_{i-1} \left( p_{i-1} - \omega_{i-1} s_{i-1} \right)$ 
		\State $s_i := w_i + \beta_{i-1} \left( s_{i-1} - \omega_{i-1} z_{i-1} \right)$
		\State $z_i := t_i + \beta_{i-1} \left( z_{i-1} - \omega_{i-1} v_{i-1} \right)$
		\State $q_i := r_i - \alpha_i s_i$
		\State $y_i := w_i - \alpha_i z_i$
		\State \textbf{begin reduction} $\left( q_i, y_i \right)$; $\left( y_i, y_i \right)$
		\State \textbf{computation} $v_i := A z_i$
		\State \textbf{end reduction}
		\State $\omega_i := \left( q_i, y_i \right)/\left( y_i, y_i \right)$
		\State $x_{i+1} := x_i + \alpha_{i} p_i + \omega_i q_i$
    \State $r_{i+1} := q_i - \omega_{i} y_i$
		\State $w_{i+1} := y_i - \omega_i \left(t_i - \alpha_i v_i \right)$
		\State \textbf{begin reduction} $\left( r_0, r_{i+1} \right)$; $\left( r_0, w_{i+1} \right)$; $\left( r_0, s_i \right)$; $\left( r_0, z_i \right)$
		\State \textbf{computation} $t_{i+1} := A w_{i+1}$
		\State \textbf{end reduction}
		\State $\beta_i := \left( \alpha_i / \omega_i \right)  \left( r_0, r_{i+1} \right) / \left( r_0, r_i \right)$
		\State $\alpha_{i+1} := \left( r_0, r_{i+1} \right) / \left( \left( r_0, w_{i+1} \right) + \beta_i \left( r_0, s_i \right) - \beta_i \omega_i \left( r_0, z_i \right) \right)$
    \EndFor
    \EndFunction
  \end{algorithmic}
\end{algorithm}

Following Step 1, the modified BiCGStab Algorithm \ref{algo::bicgstab2} now features two \textsc{spmv} operations (lines 6 and 13), which are each followed by a global reduction phase (lines 9 and 14). The first \textsc{spmv}-global reduction pair (lines 6 and 9) is separated by intermediate \textsc{axpy} operations. Starting from this algorithm, we now aim at formulating a communication-hiding version of BiCGStab by moving the \textsc{spmv} operations below the global reduction phases.

The \textsc{spmv} $z_i = A s_i$ in Algorithm \ref{algo::bicgstab2} (line 6) is rewritten using the recurrence \eqref{eq:s_i} for $s_i$:
\begin{align}
	z_i = A s_i  &= A \left( w_i + \beta_{i-1} \left( s_{i-1} - \omega_{i-1} z_{i-1} \right) \right) \notag \\
								&= t_i + \beta_{i-1} \left( z_{i-1} - \omega_{i-1} v_{i-1} \right).
\end{align}
Here we use that $z_{i-1} = A s_{i-1}$, and two new auxiliary variables are defined as the \textsc{spmv}s $t_i = A w_i$ and $v_i = A z_i$. In a similar way the second \textsc{spmv} $w_{i+1} = A r_{i+1}$ can be rewritten as a recurrence using the recursive definition of $r_{i+1}$, i.e., 
\begin{equation}
r_{i+1} = q_i - \omega_i y_i = r_i - \alpha_i s_i - \omega_i \left( w_i - \alpha_i z_i \right),
\end{equation}
see lines 7, 8 and 12. From this expression we obtain the following recurrence for $w_{i+1}$:
\begin{align}
	w_{i+1} = A r_{i+1}  &= A \left( q_i - \omega_i \left( w_i - \alpha_i z_i \right) \right) \notag \\
												&= y_i - \omega_i (t_i - \alpha_i v_i), 
\end{align}
where we use the definitions $w_i = A r_i$, $z_i = A s_i$, $t_i = A w_i$ and $v_i = A z_i$. Next, the dot-products are reformulated such that they are independent of the corresponding newly defined \textsc{spmv}s. For the first global reduction on line 9 in Alg.~\ref{algo::bicgstab2}, this implies $q_i$ and $y_i$ should be independent of $v_i$, which is clearly the case, and for the second global reduction on line 14, the variables $r_{i+1}$, $w_{i+1}$, $s_i$ and $z_i$ are required to be independent of $t_{i+1}$, which is also trivially satisfied. Hence, the \textsc{spmv} $v_i = Az_i$ can be computed below the first global reduction phase (line 9), and the second new \textsc{spmv} $t_{i+1} = A w_{i+1}$ is moved below the corresponding second global reduction (line 14). This results in the pipelined BiCGStab method shown in Algorithm \ref{algo::bicgstab3}.

Mathematically, i.e., in exact arithmetic, Algorithm \ref{algo::bicgstab3} is equivalent to standard BiCGStab. Moreover, similar to Alg.~\ref{algo::bicgstab2}, the pipelined BiCGStab Alg.~\ref{algo::bicgstab3} features only two global communication phases, yet it additionally allows for a communication-hiding strategy. After local computation of the dot-product contributions has been executed on each worker, each global reduction (lines 9 and 16) can be overlapped with the computation of the corresponding \textsc{spmv} (lines 10 and 17). This overlap hides (part of) the communication latency behind the \textsc{spmv} computations. In the remainder of this work Algorithm \ref{algo::bicgstab3} shall be referred to as unpreconditioned pipelined BiCGStab.

In finite precision arithmetic, the pipelined BiCGStab method may display different convergence behavior compared to standard BiCGStab due to the different way rounding errors are handled by the algorithm. Notably, pipelined BiCGStab features 8 lines of recursive vector operations, whereas traditional BiCGStab has only 4 lines with vector recurrences. The authors refer to the extensive literature \cite{demmel1997applied,greenbaum1989behavior,greenbaum1997estimating,gutknecht2000accuracy,meurant2006lanczos,paige1976error,paige1980accuracy,strakovs2002error,tong2000analysis} and their own related work \cite{cools2016rounding} for a more elaborate discussion on the relation between the number of \textsc{axpy}s and the propagation of local rounding errors throughout various variants of Krylov subspace algorithms. 
Hence, the p-BiCGStab residuals may deviate slightly from their original BiCGStab counterparts. Moreover, the maximal attainable accuracy of the p-BiCGStab method may be affected by the reordering of the algorithm. This is illustrated in Section \ref{sec:numerical}, where countermeasures to the accuracy loss are proposed.

\subsection{Pipelined BiCGStab vs.~Improved BiCGStab} \label{sec:ibicgstab}

\begin{table}[t]
\centering
\vspace{1.0cm}
\footnotesize
\begin{tabular}{| l | c | c | c | c | c |}
\hline 
			& \textsc{glred} & \textsc{spmv} & Flops (\textsc{axpy} + \textsc{dot-prod}) & Time (\textsc{glred} + \textsc{spmv}) & Memory \\
\hline 
BiCGStab	 	& 3 & 2 & 20 & 3 \textsc{glred} + 2 \textsc{spmv} & 7 \\
IBiCGStab		& 1 & 2 & 30 & 1 \textsc{glred} + 2 \textsc{spmv} & 10 \\
p-BiCGStab	&	2 & ~$2^*$ & 38 & 2 $\max$(\textsc{glred}, \textsc{spmv}) & 11 \\
$s$-step CA-BiCGStab & 1/$s$ & 4 & 32$s$+45 & 1/$s$ \textsc{glred} + 4 \textsc{spmv} & 4$s$+5 \\
\hline
\end{tabular}
\caption{Specifications of different unpreconditioned BiCGStab variations. 
Columns \textsc{glred} and \textsc{spmv} list the number of global reduction phases and \textsc{spmv}s per iteration respectively. 
The $^*$-symbol indicates that \textsc{spmv}s are overlapped with global reductions. 
Column \emph{Flops} shows the number of flops ($\times N$) required to compute \textsc{axpy}s and dot-products. 
The \emph{Time} column has the time spent in global all-reduce communications (\textsc{glred}s) and \textsc{spmv}s. 
\emph{Memory} counts the total number of vectors that need to be kept in memory.}
\label{tab:ibicgstab}
\end{table}

The derivation of the pipelined BiCGStab method, Algorithm \ref{algo::bicgstab3}, is essentially but one of several possible ways to obtain a more efficient parallel variant of the BiCGStab algorithm. By performing the recursive substitutions and re-orderings of the algorithm in a different manner, other parallel algorithms that are mathematically equivalent to standard BiCGStab but feature improved global communication properties may be derived.
In particular, it is possible to reduce the number of global reductions even further after obtaining Alg.~\ref{algo::bicgstab2} in Section \ref{sec:avoiding}. This is achieved by merging the reduction required to compute $\omega_i$ (line 9) with the global reduction for $\alpha_{i+1}$ and $\beta_i$ (line 14). The resulting algorithm comprises only one global reduction step per iteration. This algorithm has been proposed by Yang et al.~in 2002 as the Improved BiCGStab method (IBiCGStab) \cite{yang2002improved}. Since only one global reduction step (avoiding) but no \textsc{spmv} overlap (hiding) is performed, the IBiCGStab method performs well on setups with a heavy communication-to-computation ratio. 

We refer to Table \ref{tab:ibicgstab} for a comparison between the unpreconditioned IBiCGStab and p-BiCGStab algorithms. The \emph{Flops} and \emph{Memory} requirements for both methods are largely comparable. Only the two main time-consuming components (\textsc{glred}s and \textsc{spmv}s) are taken into account for the \emph{Time} estimates; the \textsc{axpy} operations are neglected. In the ideal scenario for pipelining where the \textsc{spmv} computation perfectly overlaps the communication time of one global reduction, the IBiCGStab method achieves a speed-up of approximately 5/3 = 1.67$\times$ compared to standard BiCGStab, whereas the p-BiCGStab method is theoretically able to attain a speed-up factor of 5/2 = 2.5$\times$.  

It is possible to combine the communication-avoiding strategy of the IBiCGStab method with a communication-hiding strategy. Starting from the IBiCGStab algorithm, additional recurrences for auxiliary variables are introduced to merge all \textsc{spmv}s into one block that directly follows the global reduction. This results in a hybrid pipelined IBiCGStab algorithm with only one global all-reduce communication phase, overlapped by \textsc{spmv}s, per iteration. 
However, to reorganize the algorithm to this state, an additional \textsc{spmv} is introduced, increasing the total number of \textsc{spmv}s per iteration from two to three. This is undesirable; the reduction of global communication latency should not come at the expense of an increased computational cost. 
More importantly, a large number of auxiliary \textsc{axpy}s is required, which threatens the stability of the algorithm and leading to very low attainable accuracy. We therefore do not expound on the details of this method. 

\subsection{Pipelined BiCGStab vs.~$s$-step CA-BiCGStab} \label{sec:sstep-bicgstab}

Algorithm \ref{algo::bicgstab2} is denoted `CA-BiCGStab' in this work to distinguish it from the standard BiCGStab Alg.~\ref{algo::bicgstab1}, which features more global reductions, and from the pipelined Alg.~\ref{algo::bicgstab3}, which overlaps global reductions with computational work. Despite the nominal similarity, Alg.~\ref{algo::bicgstab2} is unrelated to the $s$-step CA-BiCGStab algorithm proposed by Carson et al.~in \cite{carson2015communication,carson2013avoiding}. The latter algorithm is not obtained by reorganizing subsequent algorithmic operations to reduce the number of global reductions like Alg.~\ref{algo::bicgstab2}. Instead, $s$-step CA-BiCGStab simultaneously orthogonalizes $s$ Krylov subspace basis vectors prior to every $s$ iterations, such that the time spent in global communication is roughly divided by a factor of $s$. 

Table \ref{tab:ibicgstab} lists specifications of the $s$-step CA-BiCGStab algorithm from \cite{carson2015communication}, p.54, Algorithm 12. The $s$-step algorithm is very efficient for problems where the time spent by the global reduction phase significantly exceeds the time of computing an \textsc{spmv}, as is shown by the \emph{Time} estimates. 
The p-BiCGStab method, on the other hand, focuses on overlapping communication and computations, and thus benefits most from an approximately equal execution time for global reductions and \textsc{spmv}s. 
Of particular interest in this context is the work on pipelined methods with a pipeline length of $l$ iterations, presented for GMRES in \cite{ghysels2013hiding}, which allows to overlap the global reduction by multiple iterations. Currently no $l$-length version of the pipelined BiCGStab method is available. We aim to address this non-trivial question in future research. Note that for $s$-step methods, the inclusion of a preconditioner is generally challenging, whereas it is (theoretically) straightforward in the case of pipelined methods. This is illustrated for the p-BiCGStab method in Section \ref{sec:precpipebicgstab}.

Compared to other BiCGStab variants, the $s$-step CA-BiCGStab method requires a larger average number of \textsc{spmv}s per iteration and significantly more flops to compute the growing number of \textsc{axpy}s and dot-products. The large number of extra operations affects the stability of the method for increasing values of $s$, see the analysis in \cite{carson2015communication}. The latter phenomenon is also observed for pipelined methods, and is treated in more detail for p-BiCGStab in Section \ref{sec:numerical} of this work.

\subsection{Preconditioned pipelined BiCGStab}  \label{sec:precpipebicgstab}

The derivation of the pipelined BiCGStab algorithm can easily be extended to the preconditioned BiCGStab method shown in Algorithm \ref{algo::bicgstab4}. 
A right-preconditioned system $AM^{-1}y = b$ with $M x = y$ is considered, where $M$ is the preconditioning operator.
The preconditioned algorithm differs only slightly from Alg.~\ref{algo::bicgstab1}: lines 4 and 9 are added to implement the preconditioner, and the solution update on line 13 now uses the preconditioned direction vectors; the remaining expressions in Alg.~\ref{algo::bicgstab4} are identical to the ones in Alg.~\ref{algo::bicgstab1}.
The preconditioned variables are denoted by their respective variable with a hat-symbol, e.g.~$\hat{p}_i = M^{-1} p_i$ and $\hat{q}_i = M^{-1} q_i$.

\begin{algorithm}[t]
  \caption{Preconditioned BiCGStab}
  \label{algo::bicgstab4}
  \begin{algorithmic}[1]
  	\Function{prec-bicgstab}{$A$, $M^{-1}$, $b$, $x_0$}
    \State $r_0 := b - Ax_0$; $p_0 := r_0$  
    \For{$i = 0,1,2, \dots$}\
		\State \textbf{computation} $\hat{p}_i := M^{-1} p_i$
    \State \textbf{computation} $s_i := A\hat{p}_{i}$ 
		\State \textbf{begin reduction} $\left( r_0, s_i \right)$ \textbf{end reduction}
    \State $\alpha_i := \left( r_0, r_i \right) / \left( r_0, s_i \right)$ 
		\State $q_i := r_i - \alpha_i s_i$
		\State \textbf{computation} $\hat{q}_i := M^{-1} q_i$
		\State \textbf{computation} $y_i := A \hat{q}_i$
		\State \textbf{begin reduction} $\left( q_i, y_i \right)$; $\left( y_i, y_i \right)$ \textbf{end reduction}
		\State $\omega_i := \left( q_i, y_i \right)/\left( y_i, y_i \right)$
    \State $x_{i+1} := x_i + \alpha_{i} \hat{p}_i + \omega_i \hat{q}_i$
    \State $r_{i+1} := q_i - \omega_{i} y_i$
		\State \textbf{begin reduction} $\left( r_0, r_{i+1} \right)$ \textbf{end reduction}
    \State $\beta_i := \left( \alpha_i / \omega_i \right)  \left( r_0, r_{i+1} \right) / \left( r_0, r_i \right)$
    \State $p_{i+1} := r_{i+1} + \beta_i \left( p_i - \omega_i s_i \right)$ 
    \EndFor
    \EndFunction
  \end{algorithmic}
\end{algorithm}

The preconditioned pipelined BiCGStab method is derived starting from Alg.~\ref{algo::bicgstab4} following a framework that is comparable to the derivation of unpreconditioned pipelined algorithm. However, in this case the preconditioned variables $\hat{p}_i$ (line 4) and $\hat{q}_i$ (line 9) are also reformulated as recurrences where required, similar to the reformulation of the \textsc{spmv}s for $s_i$ (line 5) and $y_i$ (line 10).
First, the number of global communication phases is reduced by joining the global reductions required to compute $\alpha_i$ and $\beta_i$ together into one global reduction phase. To this aim, we rewrite the variables $\hat{p}_i$ and $s_i$ in the first preconditioned \textsc{spmv} (line 4-5) as recurrences, i.e.,
\begin{align} \label{eq:hatp_i}
	\hat{p}_i = M^{-1} p_i	&= M^{-1} \left( r_i + \beta_{i-1} \left( p_{i-1} - \omega_{i-1} s_{i-1} \right) \right) \notag \\
													&= \hat{r}_i + \beta_{i-1} \left( \hat{p}_{i-1} - \omega_{i-1} \hat{s}_{i-1} \right),
\end{align}
where the new variables $\hat{r}_i := M^{-1} r_i$ and $\hat{s}_i := M^{-1} s_i$ are defined, and furthermore we write
\begin{align} \label{eq:s_i2}
	s_i = A \hat{p}_i 	&= A \left( \hat{r}_i + \beta_{i-1} \left( \hat{p}_{i-1} - \omega_{i-1} \hat{s}_{i-1} \right) \right) \notag \\
											&= w_i + \beta_{i-1} \left( s_{i-1} - \omega_{i-1} z_{i-1} \right),
\end{align}
where $\hat{p}_i$ is substituted by the recurrence \eqref{eq:hatp_i}, and the variables $w_i = A \hat{r}_i = A M^{-1} r_i$ and $z_i = A \hat{s}_i = A M^{-1} s_i$ are introduced. 
Furthermore, using the definitions above, the variables $\hat{q}_i$ and $y_i$ in the second preconditioned \textsc{spmv} (line 9-10) are rewritten as recurrence relations, yielding for $\hat{q}_i = M^{-1} q_i$
\begin{equation}
	\hat{q}_i = M^{-1} q_i = M^{-1} \left( r_i - \alpha_i s_i \right) = \hat{r}_i - \alpha_i \hat{s}_i,
\end{equation}
and for the variable $y_i = A \hat{q}_i$ the recurrence
\begin{equation} \label{eq:y_i2}
	y_i = A \hat{q}_i = A \left( \hat{r}_i - \alpha_i \hat{s}_i \right) = w_i - \alpha_i z_i.
\end{equation}
Observe how the recurrences \eqref{eq:s_i2} and \eqref{eq:y_i2} for $s_i$ and $y_i$ respectively are identical to their counterparts \eqref{eq:s_i} and \eqref{eq:y_i} in the unpreconditioned case. Consequently, definition \eqref{eq:alpha_i} for $\alpha_i$ can again be used here. This expression does not depend on the variables $p_{i+1}$, $\hat{p}_i$ and $s_i$. The global reduction for $\alpha_i$ can thus be moved upward and merged with the global reduction for $\beta_i$. Hence, the number of global reductions is reduced from three to two at the cost of two additional 
recursive vector operations, resulting in a communication-avoiding preconditioned BiCGStab algorithm.

\begin{algorithm}[t]
  \caption{Preconditioned pipelined BiCGStab}
  \label{algo::bicgstab5}
  \begin{algorithmic}[1]
  	\Function{prec-p-bicgstab}{$A$, $M^{-1}$, $b$, $x_0$}
    \State $r_0 := b - Ax_0$; $\hat{r}_0 := M^{-1} r_0$; $w_0 := A \hat{r}_0$; $\hat{w}_0 := M^{-1} w_0$
		\State $t_0 := A \hat{w}_0$; $\alpha_0 := \left( r_0, r_0 \right) / \left( r_0, w_0 \right)$; $\beta_{-1} := 0$  
    \For{$i = 0,1,2, \dots$}
    \State $\hat{p}_i := \hat{r}_i + \beta_{i-1} \left( \hat{p}_{i-1} - \omega_{i-1} \hat{s}_{i-1} \right)$ 
		\State $s_i := w_i + \beta_{i-1} \left( s_{i-1} - \omega_{i-1} z_{i-1} \right)$
		\State $\hat{s}_i := \hat{w}_i + \beta_{i-1} \left( \hat{s}_{i-1} - \omega_{i-1} \hat{z}_{i-1} \right)$
		\State $z_i := t_i + \beta_{i-1} \left( z_{i-1} - \omega_{i-1} v_{i-1} \right)$
		\State $q_i := r_i - \alpha_i s_i$
		\State $\hat{q}_i := \hat{r}_i - \alpha_i \hat{s}_i$
		\State $y_i := w_i - \alpha_i z_i$
		\State \textbf{begin reduction} $\left( q_i, y_i \right)$; $\left( y_i, y_i \right)$
		\State \textbf{computation} $\hat{z}_i := M^{-1} z_i$
		\State \textbf{computation} $v_i := A \hat{z}_i$
		\State \textbf{end reduction}
		\State $\omega_i := \left( q_i, y_i \right)/\left( y_i, y_i \right)$
		\State $x_{i+1} := x_i + \alpha_{i} \hat{p}_i + \omega_i \hat{q}_i$
    \State $r_{i+1} := q_i - \omega_{i} y_i$
		\State $\hat{r}_{i+1} := \hat{q}_i - \omega_i \left( \hat{w}_i - \alpha_i \hat{z}_i \right)$
		\State $w_{i+1} := y_i - \omega_i \left( t_i - \alpha_i v_i \right)$
		\State \textbf{begin reduction} $\left( r_0, r_{i+1} \right)$; $\left( r_0, w_{i+1} \right)$; $\left( r_0, s_i \right)$; $\left( r_0, z_i \right)$
		\State \textbf{computation} $\hat{w}_{i+1} := M^{-1} w_{i+1}$
		\State \textbf{computation} $t_{i+1} := A \hat{w}_{i+1}$
		\State \textbf{end reduction}
		\State $\beta_i := \left( \alpha_i / \omega_i \right)  \left( r_0, r_{i+1} \right) / \left( r_0, r_i \right)$
		\State $\alpha_{i+1} := \left( r_0, r_{i+1} \right) / \left(  \left( r_0, w_{i+1} \right) + \beta_i \left( r_0, s_i \right) - \beta_i \omega_i \left( r_0, z_i \right) \right)$
    \EndFor
    \EndFunction
  \end{algorithmic}
\end{algorithm}

We then focus on hiding the global communication phases behind the preconditioned \textsc{spmv}s, which will allow to overlap both the preconditioner application and the \textsc{spmv} with the global reduction. The two preconditioned \textsc{spmv}s that compute $\hat{s}_i = M^{-1} s_i$, $z_i = A \hat{s}_i$ and $\hat{r}_{i+1} = M^{-1} r_{i+1}$, $w_{i+1} = A \hat{r}_{i+1}$ are rewritten, such that the \textsc{spmv}s can be placed below the corresponding global reduction phases. This is achieved by introducing recurrences and auxiliary variables as follows:
\begin{align}
	\hat{s}_i = M^{-1} s_i &= M^{-1} \left( w_i + \beta_{i-1} \left( s_{i-1} - \omega_{i-1} z_{i-1} \right) \right) \notag \\
													&= \hat{w}_i + \beta_{i-1} \left( \hat{s}_{i-1} - \omega_{i-1} \hat{z}_{i-1} \right),
\end{align}
where $\hat{w}_i = M^{-1} w_i$ and $\hat{z}_i = M^{-1} z_i$ are introduced, and
\begin{align}
	z_i = A \hat{s}_i 	&= A \left( \hat{w}_i + \beta_{i-1} \left( \hat{s}_{i-1} - \omega_{i-1} \hat{z}_{i-1} \right) \right) \notag \\
											&= t_i + \beta_{i-1} \left( z_{i-1} - \omega_{i-1} v_{i-1} \right),
\end{align}
where $t_i = A \hat{w}_i$ and $v_i = A \hat{z}_i$ are defined. Additionally, using these new definitions, the following recurrences for $\hat{r}_{i+1}$ and $w_{i+1}$ can be derived
\begin{align}
	\hat{r}_{i+1} = M^{-1} r_{i+1} &= M^{-1} \left( q_i - \omega_i \left( w_i - \alpha_i z_i \right) \right) \notag \\
																	&= \hat{q}_i - \omega_i \left( \hat{w}_i - \alpha_i \hat{z}_i \right),
\end{align}
and 
\begin{align}
		w_{i+1} = A \hat{r}_{i+1} 	&= A \left( \hat{q}_i - \omega_i \left( \hat{w}_i - \alpha_i \hat{z}_i \right) \right) \notag \\
																&= y_i - \omega_i \left( t_i - \alpha_i v_i \right).
\end{align}
After moving the preconditioned \textsc{spmv}s $\hat{w}_i = M^{-1} w_i$, $t_i = A \hat{w}_i$ and $\hat{z}_i = M^{-1} z_i$, $v_i = A \hat{z}_i$ 
below the dot-product computations of which they are independent, and following a minor reordering of operations, 
the above recurrences result in Algorithm \ref{algo::bicgstab5}. This algorithm is denoted as preconditioned pipelined BiCGStab.

Note that Alg.~\ref{algo::bicgstab5} can alternatively be derived directly from the unpreconditioned pipelined BiCGStab Alg.~\ref{algo::bicgstab3} by adding the proper preconditioned variables. This includes recursively computing $\hat{p}_i = M^{-1} p_i$ (line 5), $\hat{s}_i = M^{-1} s_i$ (line 7), $\hat{q}_i = M^{-1} q_i$ (line 10) and $\hat{r}_{i+1} = M^{-1} r_{i+1}$ (line 19), whose recurrences can be derived directly from the corresponding unpreconditioned variables, and defining the preconditioned variables $\hat{z}_i = M^{-1} z_i$ (line 13) and $\hat{w}_{i+1} = M^{-1} w_{i+1}$ (line 22), which are computed explicitly right before their respective \textsc{spmv}s.

Similar to the unpreconditioned case, the preconditioned pipelined BiCGStab Alg.~\ref{algo::bicgstab5} allows for the overlap of the two \textsc{spmv} operations (lines 14 and 23) with the global reduction phases (lines 12 and 20). In addition, due to the consecutive application of the preconditioner $M^{-1}$ (lines 13 and 22) and the operator $A$, the global reductions also overlap with preconditioning. Depending on the preconditioner choice, this possibly results in a much better hiding of global communication when the application of $A$ is not sufficient to cover the global reduction time frame. For good parallel performance in practice the preconditioner application should be compute-bound, requiring only limited communication between neighboring processors.

\section{Numerical results} \label{sec:numerical} 

In this section a variety of numerical experiments is reported to benchmark the convergence of the pipelined BiCGStab method. 
Results comparing BiCGStab and p-BiCGStab on a wide range of matrices from applications are presented in Section \ref{sec:matrix}.
A residual replacement strategy is presented in Section \ref{sec:replacement} to increase the maximal attainable accuracy and robustness of the pipelined method.

\begin{table}[t]
\centering
\vspace{1.0cm}
\scriptsize
\begin{tabular}{| l | r r r r r | r r | r r |}
\hline 
 Matrix & Prec  & $\kappa(A)$ & $N$ & \#nnz & $\|r_0\|_2$ & \multicolumn{2}{|c|}{BiCGStab} & \multicolumn{2}{|c|}{p-BiCGStab} \\
 & & & & & & iter & $\|b-A x_i\|_2$  & iter & $\|b-A x_i\|_2$  \\
\hline 
 1138\_bus& ILU  & 8.6e+06  & 1138   & 4054     & 4.3e+01  & 	89 		 & 1.4e-05	& 95 	& 1.8e-05	\\
 add32    & ILU  & 1.4e+02  & 4960   & 19,848   & 8.0e-03  &  19     & 5.9e-09	& 19  & 5.9e-09 \\
 bcsstk14 & ILU  & 1.3e+10  & 1806   & 63,454   & 2.1e+09  & 	315    & 1.6e+03	& 322 & 1.2e+03 \\
 bcsstk18 & ILU  & 6.5e+01  & 11,948 & 149,090  & 2.6e+09  & 	84     & 2.2e+03	& 102 & 2.0e+03 \\
 bcsstk26 & ILU  & 1.7e+08  & 1922   & 30,336   & 3.5e+09  & 	113    & 2.8e+03	& 107 & 2.9e+03 \\
 bcsstm25 &   -  & 6.1e+09  & 15,439 & 15,439   & 6.9e+07  & 	928    & 6.8e+01	& 825 & 6.5e+01 \\
 bfw782a  & ILU  & 1.7e+03  & 782    & 7514     & 3.2e-01  & 	72     & 1.1e-07	& 65  & 6.2e-08 \\
 bwm2000  &   -  & 2.4e+05  & 2000   & 7996     & 1.1e+03  & 	1156   & 6.6e-04	& 1162& 9.1e-04 \\
 cdde6    & ILU  & 1.8e+02  & 961    & 4681     & 5.8e-01  & 	9      & 2.2e-07	& 9   & 2.2e-07 \\
 fidap014 &   -  & 3.5e+16  & 3251   & 65,747   & 2.7e+06  & 	121    & 2.6e+00	& 123 & 2.6e+00 \\
 fs\_760\_3& ILU & 1.0e+20  & 760    & 5816     & 1.6e+07  & 	930    & 1.4e+01	& 709 & 1.1e+01 \\
 jagmesh9 &   -  & 6.0e+03  & 1349   & 9101     & 6.8e+00  &  1022   & 6.4e-06	& 996 & 6.6e-06 \\
 jpwh\_991& ILU  & 1.4e+02  & 991    & 6027     & 3.8e-01  &  9      & 2.9e-07	& 9   & 2.9e-07 \\
 orsreg\_1& ILU  & 6.7e+03  & 2205   & 14,133   & 4.8e+00  &  25     & 2.7e-06	& 25  & 2.7e-06 \\
 pde2961  & ILU  & 6.4e+02  & 2961   & 14,585   & 2.9e-01  & 	31     & 1.3e-07	& 31  & 4.5e-08 \\
 rdb3200l &   -  & 1.1e+03  & 3200   & 18,880   & 1.0e+01  & 	149    & 5.1e-06	& 145 & 9.1e-06 \\
 s3dkq4m2 &   -  & 1.9e+11  & 90,449 & 2,455,670& 6.8e+01  & 	3736   & 5.8e-05	& 3500& 6.2e-05 \\
 saylr4   & ILU  & 6.9e+06  & 3564   & 22,316   & 3.1e-03  & 	40     & 3.0e-09	& 39  & 1.5e-09 \\
 sherman3 & ILU  & 5.5e+18  & 5005   & 20,033   & 1.8e+01  & 	98     & 3.7e-06	& 83  & 9.1e-06 \\
 sstmodel &   -  & 2.7e+18  & 3345   & 22,749   & 7.9e+00  & 	6968   & 7.7e-06	& 4399& 7.5e-06 \\
 utm5940  & ILU  & 4.3e+08  & 5940   & 83,842   & 3.6e-01  & 	223    & 4.0e-08	& 244 & 3.5e-07 \\
\hline
 \multicolumn{6}{|l|}{Average iter deviation wrt BiCGStab} & & & -3.5\% & \\
\hline
\end{tabular}
\caption{Collection of matrices from Matrix Market with condition number $\kappa(A)$, number of rows/columns $N$ and total number of nonzeros \#nnz. A linear system with right-hand side $b = A\hat{x}$ where $\hat{x}_i = 1/\sqrt{N}$ is solved with each of these matrices with the presented BiCGStab and pipelined BiCGStab algorithms. The initial guess is all-zero $x_0 = 0$. An Incomplete LU preconditioner with zero fill-in (ILU0) is included where applicable. The number of iterations required to reach a tolerance of 1e-6 on the scaled residual, i.e., $\|r_i\|_2 / \|r_0\|_2 \leq 10^{-6}$, is shown along with the corresponding true residual norm $\|b-A x_i\|_2$.
}
\label{tab:matrix_market}
\end{table}

\subsection{Pipelined BiCGStab benchmark on Matrix Market problems} \label{sec:matrix}

Numerical results on a wide range of different linear systems are presented. 
Table \ref{tab:matrix_market} lists a collection of test matrices from Matrix Market\footnote{\url{http://math.nist.gov/MatrixMarket/}}, 
with their respective condition number $\kappa$, number of rows $N$ and total number of nonzero elements \#nnz. A linear system with exact solution 
$\hat{x}_j = 1/\sqrt{N}$ and right-hand side $b = A\hat{x}$ is solved for each of these matrices.
The system is solved using both standard BiCGStab, Alg.~\ref{algo::bicgstab4}, and the new pipelined BiCGStab variant, Alg.~\ref{algo::bicgstab5}. The initial guess is the all-zero vector $x_0 = 0$ for both methods. An Incomplete LU factorization preconditioner is included to reduce the number of Krylov iterations where applicable.
The stopping criterion defined on the scaled recursive residual is $\|r_i\|_2 / \|r_0\|_2 \leq 10^{-6}$.

\begin{figure}[t]
\begin{center}
\begin{tabular}{cc}
\includegraphics[width=0.48\textwidth]{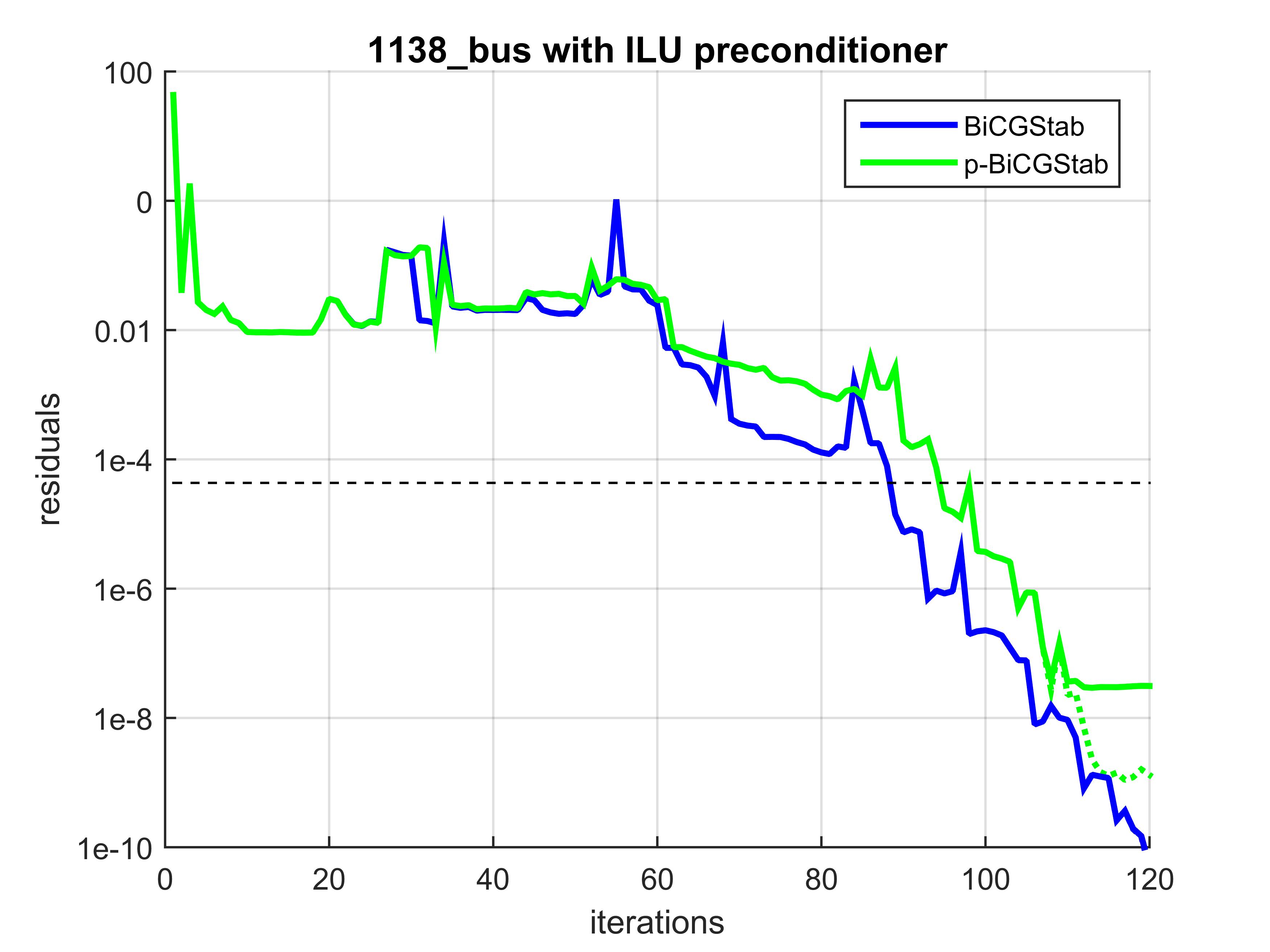} &
\includegraphics[width=0.48\textwidth]{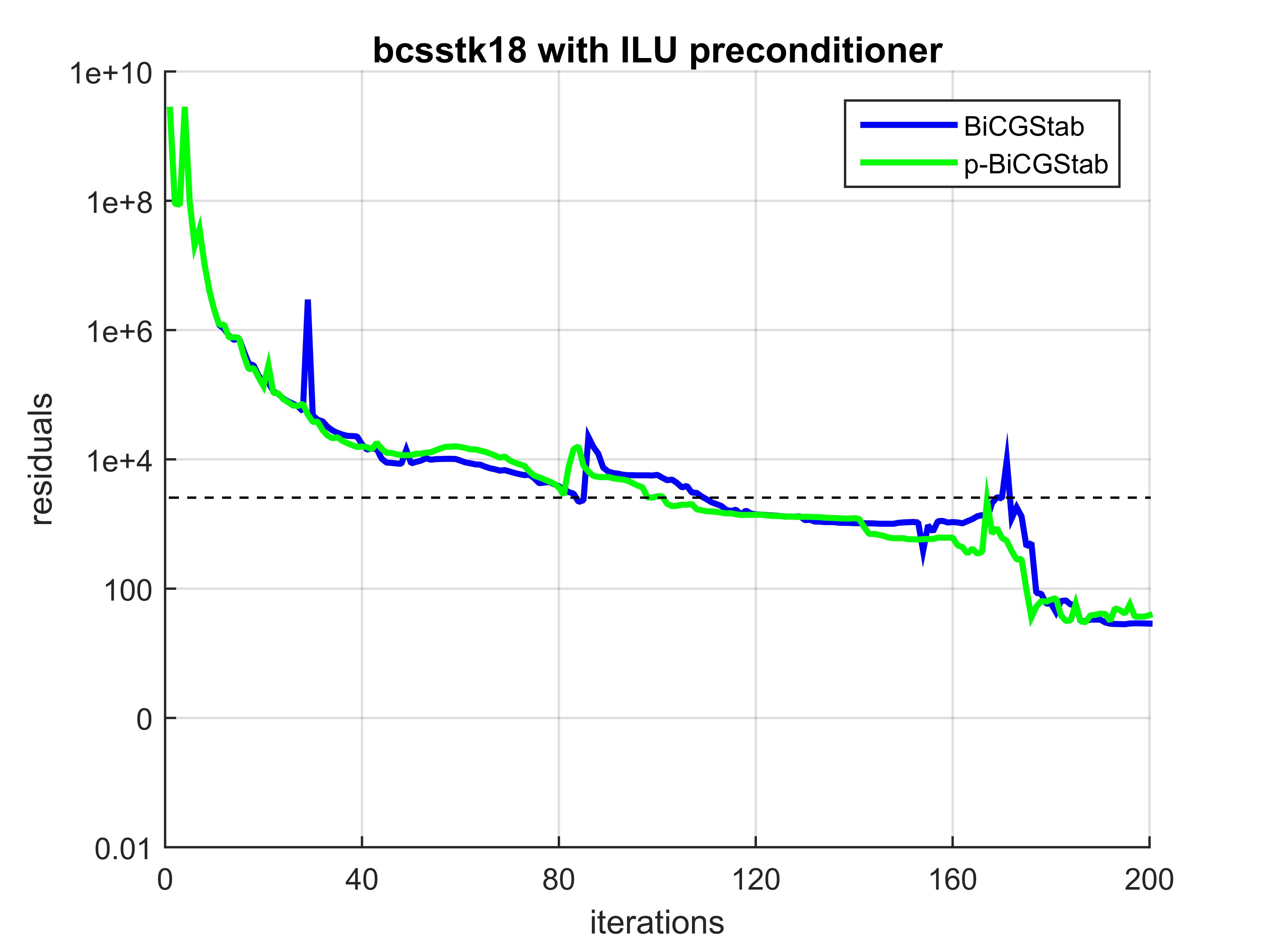} \vspace{0.2cm} \\ 
\includegraphics[width=0.48\textwidth]{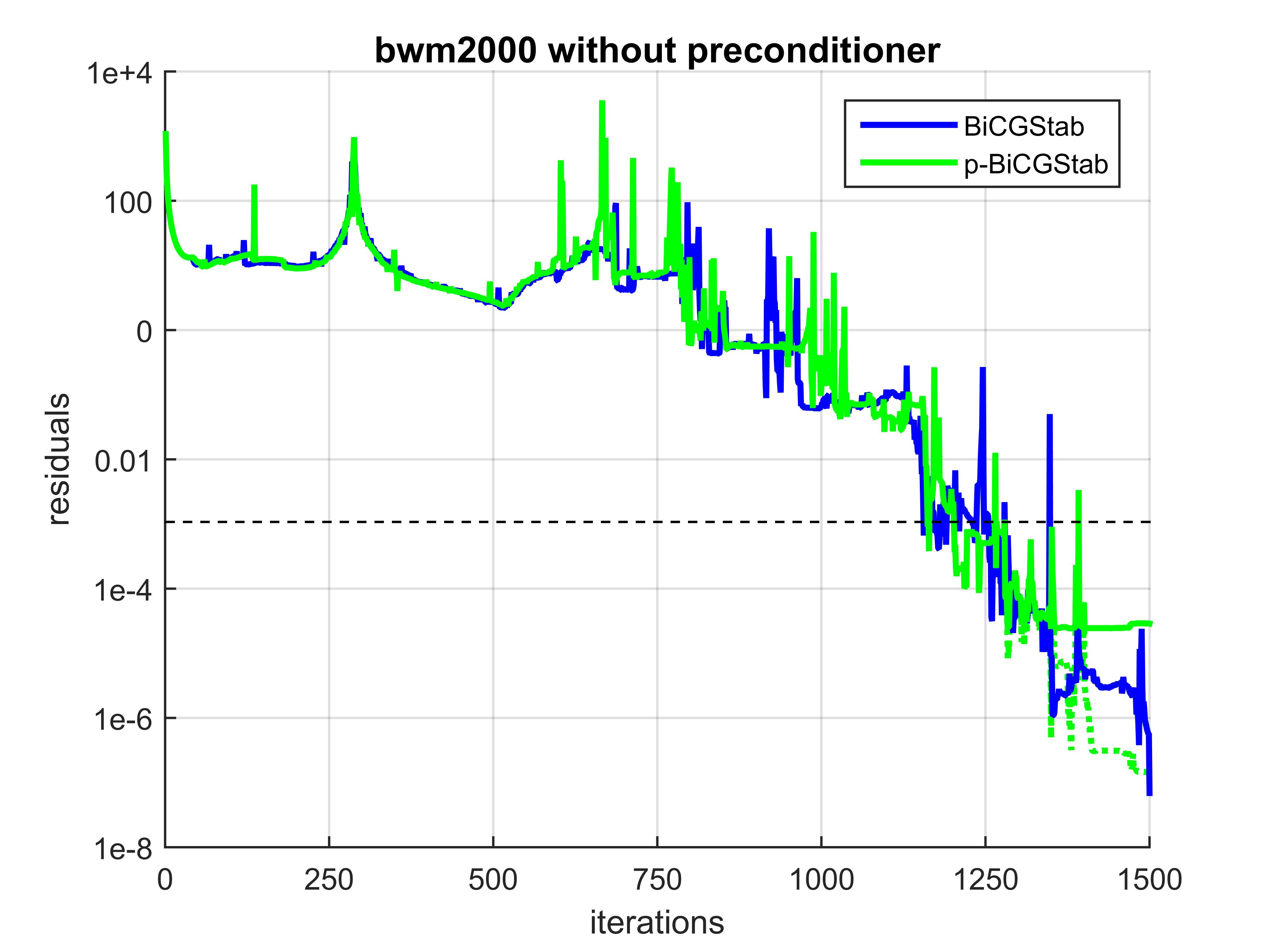} &
\includegraphics[width=0.48\textwidth]{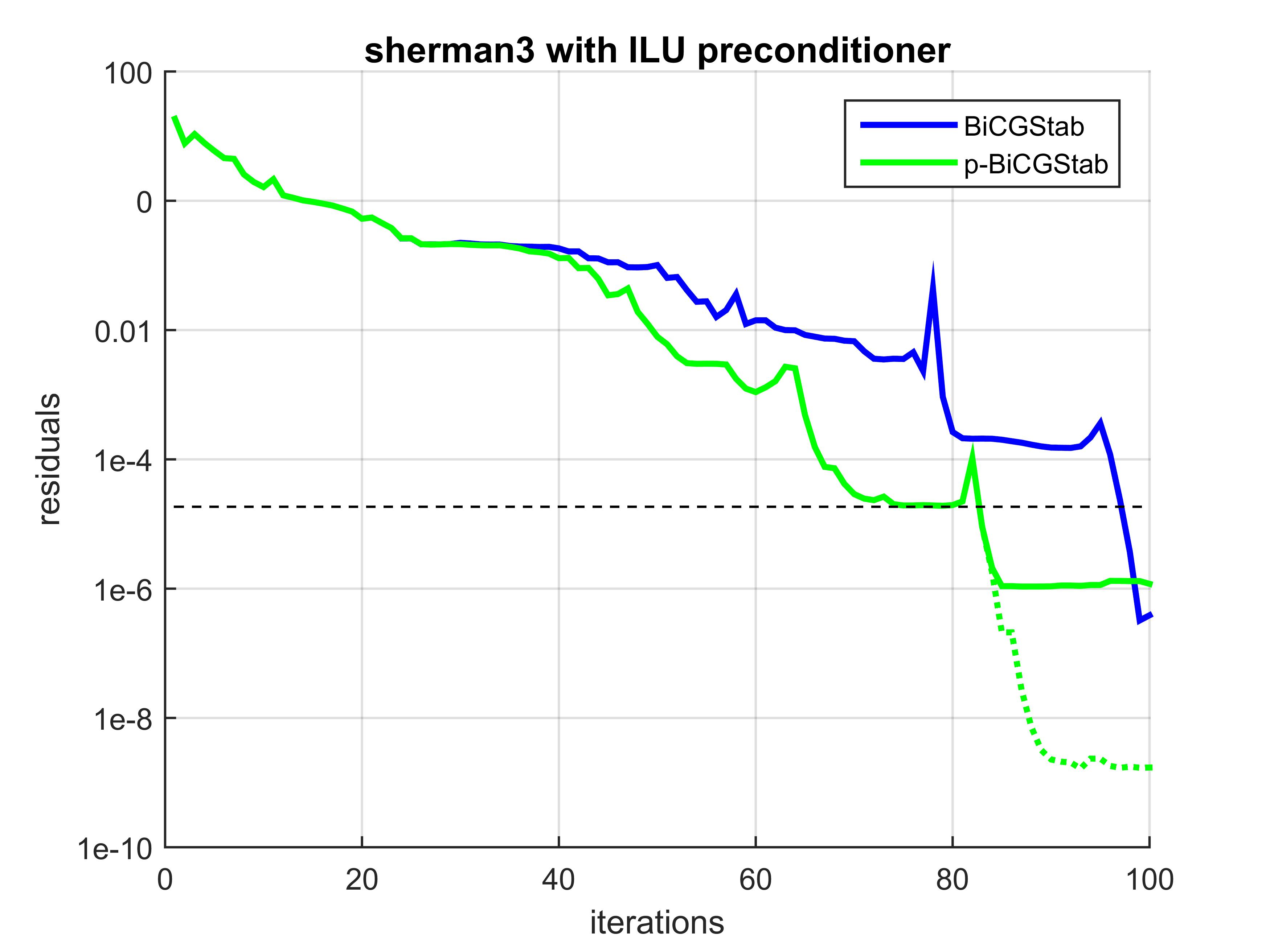}
\end{tabular}
\end{center}
\caption{Residual history of BiCGStab (blue) and p-BiCGStab (green) for different test matrices. Specifications: see Table \ref{tab:matrix_market}. 
Solid color lines represent the true residual norms $\|b-Ax_i\|_2$, while the dotted color lines are the norms of the recursive residuals $\|r_i\|_2$. The dashed line marks the tolerance 1e-6 on the scaled residual that was used as a stopping criterion.
\label{fig:conv_examples}}
\end{figure}

Table \ref{tab:matrix_market} lists the number of iterations required to reach the stopping criterion as well as the final true residual norm $\|b-A x_i\|_2$ for both methods. For most problems the preset tolerance of $10^{-6}$ on the scaled residual is achievable by p-BiCGStab in a comparable number of iterations with respect to standard BiCGStab. Averaged over all matrices, 3.5\% less iterations are required by the pipelined algorithm. For specific problems p-BiCGStab requires significantly more iterations (e.g.~\texttt{bcsstk18}, \texttt{utm5940}), whereas for others less iterations are required (e.g.~\texttt{bcsstm25}, \texttt{sstmodel}).

Fig.~\ref{fig:conv_examples} illustrates the convergence histories corresponding to a few randomly selected Matrix Market problems listed in Table \ref{tab:matrix_market}. The per-iteration residual norms for BiCGStab and p-BiCGStab on the \texttt{1138\_bus} (top-left), \texttt{bcsstk18} (top-right), \texttt{bwm2000} (bottom-left) and \texttt{sherman3} (bottom-right) matrices are shown, all with ILU0 preconditioner except the \texttt{bwm2000} problem. The tolerance of $10^{-6}$ on the scaled residual used as stopping criterion in Table \ref{tab:matrix_market} is marked by the dashed black line. Note that a stagnation of the true residual (solid colored lines) does not necessarily imply stagnation of the recursive residual (dotted colored lines). The latter may continue to decrease although the solution no longer improves with additional iterations, cf.~\cite{greenbaum1997estimating}.

Table \ref{tab:matrix_market} and Fig.~\ref{fig:conv_examples} show that the residual norm history for BiCGStab and p-BiCGStab, although comparable, is not identical. This discrepancy is due to the different way rounding errors are propagated through the algorithms in finite precision arithmetic. Indeed, although BiCGStab and p-BiCGStab are mathematically equivalent algorithms, small differences in convergence may arise as a result of rounding error propagation in finite precision arithmetic. Due to the generally non-smooth convergence history of the BiCGStab method, these effects are more pronounced for BiCGStab compared to the pipelined Conjugate Gradient method \cite{ghysels2014hiding}, where the behavior of the traditional and pipelined algorithms is largely identical. We expound on the numerical stability of the algorithm in finite precision in Section \ref{sec:replacement}.

\subsection{Improving attainable accuracy: residual replacement strategy} \label{sec:replacement}

\begin{table}[t]
\centering
\vspace{1.0cm}
\scriptsize
\begin{tabular}{| l | r | r r | r r | r r r r |}
\hline 
 Matrix & $\|r_0\|_2$ &  \multicolumn{2}{|c|}{BiCGStab} & \multicolumn{2}{|c|}{p-BiCGStab} & \multicolumn{2}{|c}{p-BiCGStab-rr} & & \\
	      &               & iter 	& $\|b-A x_i\|_2$  								 & iter	& $\|b-A x_i\|_2$ 										& iter & $\|b-A x_i\|_2$ & $k$ & \#nrr\\
\hline 
 1138\_bus& 4.3e+01  & 124  &  1.8e-11 & 130  & 4.0e-09 & 220    & 7.4e-12 &  35   & 3 \\
 add32    & 8.0e-03  & 46   &  7.8e-18 & 42   & 5.0e-16 & 51     & 5.7e-18 &  10   & 2 \\
 bcsstk14 & 2.1e+09  & 559  &  7.3e-06 & 444  & 6.6e-01 & 522    & 3.8e-03 &  200  & 2 \\
 bcsstk18 & 2.6e+09  & 523  &  4.8e-06 & 450  & 1.1e-01 & 725    & 2.8e-05 &  50   & 7 \\
 bcsstk26 & 3.5e+09  & 414  &  1.1e-05 & 216  & 5.7e-01 & 475    & 8.6e-04 &  30   & 6 \\
 bcsstm25 & 6.9e+07  & -    &  3.2e+00 & -    & 3.8e+00 & -      & 4.6e+00 &  1000 & 9 \\
 bfw782a  & 3.2e-01  & 117  &  9.5e-14 & 106  & 5.1e-13 & 133    & 2.6e-15 &  20   & 4 \\
 bwm2000  & 1.1e+03  & 1733 &  2.5e-09 & 1621 & 1.4e-05 & 2231   & 3.8e-08 &  500  & 3 \\
 cdde6    & 5.8e-01  & 151  &  8.1e-14 & 147  & 2.0e-11 & 159    & 2.1e-15 &  10   & 13\\
 fidap014 & 2.7e+06  & -    &  4.3e-03 & -    & 9.7e-03 & -      & 4.3e-03 &  50   & 3 \\
 fs\_760\_3& 1.6e+07 & 1979 &  1.2e-05 & 1039 & 5.1e-02 & 4590   & 1.1e-05 &  900  & 3 \\
 jagmesh9 & 6.8e+00  & 6230 &  2.4e-14 & 3582 & 5.8e-09 & 9751   & 1.1e-11 &  500  & 13\\
 jpwh\_991& 3.8e-01  & 53   &  1.3e-14 & 54   & 1.8e-12 & 63     & 2.5e-15 &  10   & 4 \\
 orsreg\_1& 4.8e+00  & 51   &  4.0e-11 & 52   & 3.7e-09 & 56     & 5.9e-12 &  10   & 3 \\
 pde2961  & 2.9e-01  & 50   &  4.5e-15 & 48   & 3.4e-13 & 52     & 1.4e-15 &  10   & 3 \\
 rdb3200l & 1.0e+01  & 178  &  3.7e-08 & 167  & 9.9e-08 & 181    & 3.4e-08 &  100  & 1 \\
 s3dkq4m2 & 6.8e+01  & -    &  1.0e-05 & -    & 1.4e-05 & -      & 1.3e-05 &  1000 & 9 \\
 saylr4   & 3.1e-03  & 52   &  4.0e-12 & 43   & 7.5e-11 & 46     & 1.8e-12 &  10   & 4 \\
 sherman3 & 1.8e+01  & 111  &  2.5e-11 & 100  & 2.8e-07 & 128    & 6.2e-11 &  20   & 4 \\
 sstmodel & 7.9e+00  & -    &  5.1e-06 & -    & 3.1e-06 & -      & 4.5e-06 &  1000 & 9 \\
 utm5940  & 3.6e-01  & 256  &  3.0e-12 & 248  & 4.3e-08 & 395    & 2.9e-11 &  100  & 3 \\
\hline
 \multicolumn{4}{|l|}{Average iter deviation wrt BiCGStab} & -11.0\% & & 22.1\% & & & \\
 \multicolumn{4}{|l|}{Average \#nrr wrt p-BiCGStab-rr iter} & & & & & & 2.4\% \\
\hline
\end{tabular}
\caption{Collection of matrices from Matrix Market. See Table \ref{tab:matrix_market} for additional specifications. A linear system with right-hand side $b = A\hat{x}$ where $\hat{x}_i = 1/\sqrt{N}$ is solved with each of these matrices with the BiCGStab, p-BiCGStab and p-BiCGStab-rr algorithms. The initial guess is all-zero $x_0 = 0$. An Incomplete LU preconditioner with zero fill-in (ILU0) is included where applicable, see Table \ref{tab:matrix_market}. The number of iterations required to reach the maximal attainable accuracy on the residual is shown, along with the corresponding true residual norm $\|b-A x_i\|_2$. A `-' symbol denotes failure to converge to a tolerance of 1e-8 on the scaled residual, i.e., $\|r_i\|_2/\|r_0\|_2 \leq 10^{-8}$, within 10,000 iterations. For the p-BiCGStab-rr method the table indicates the replacement period $k$ and the total number of replacement steps \#nrr. 
}
\label{tab:matrix_market2}
\end{table}

\begin{figure}[t]
\begin{center}
\begin{tabular}{cc}
\includegraphics[width=0.48\textwidth]{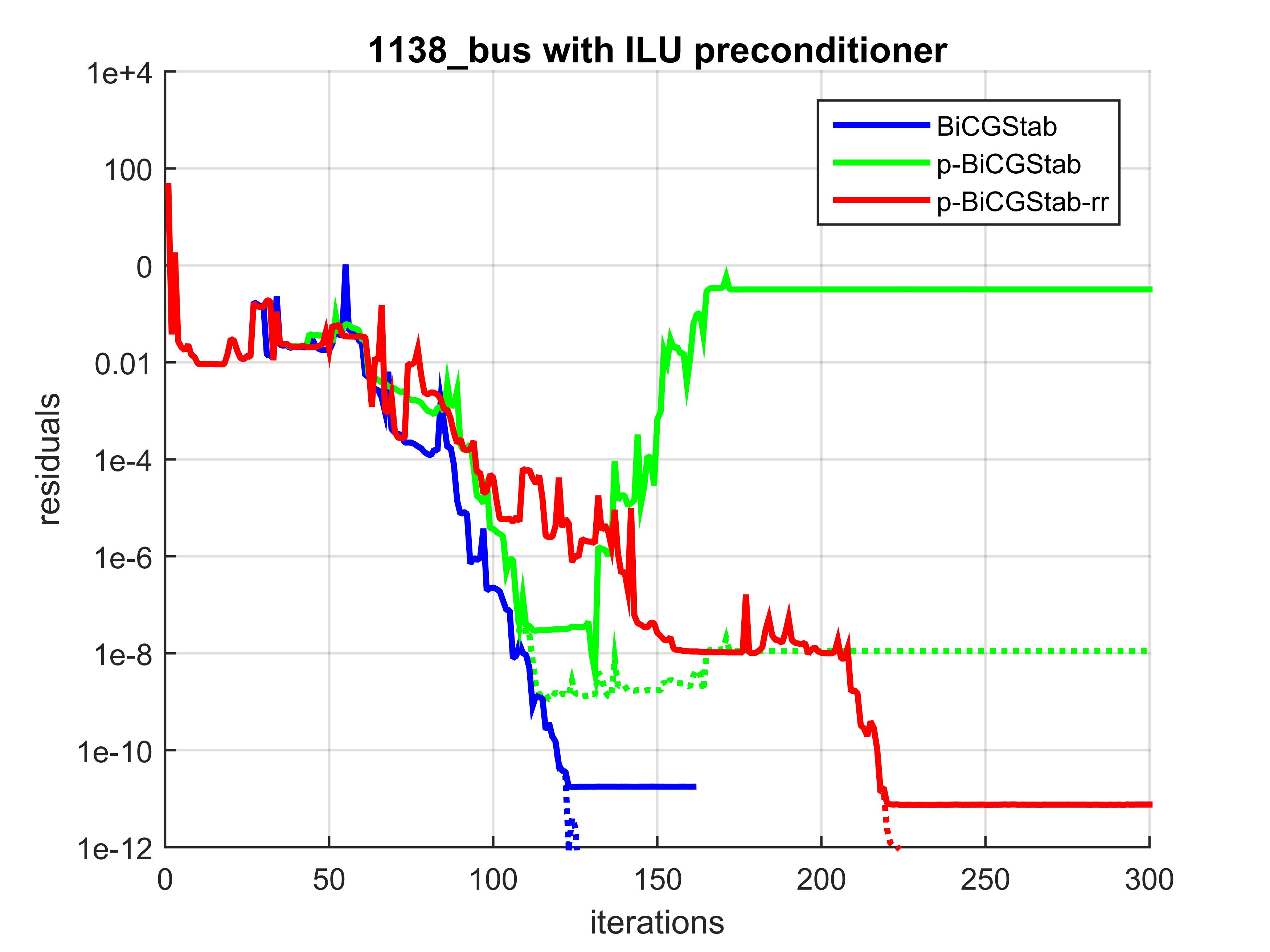} &
\includegraphics[width=0.48\textwidth]{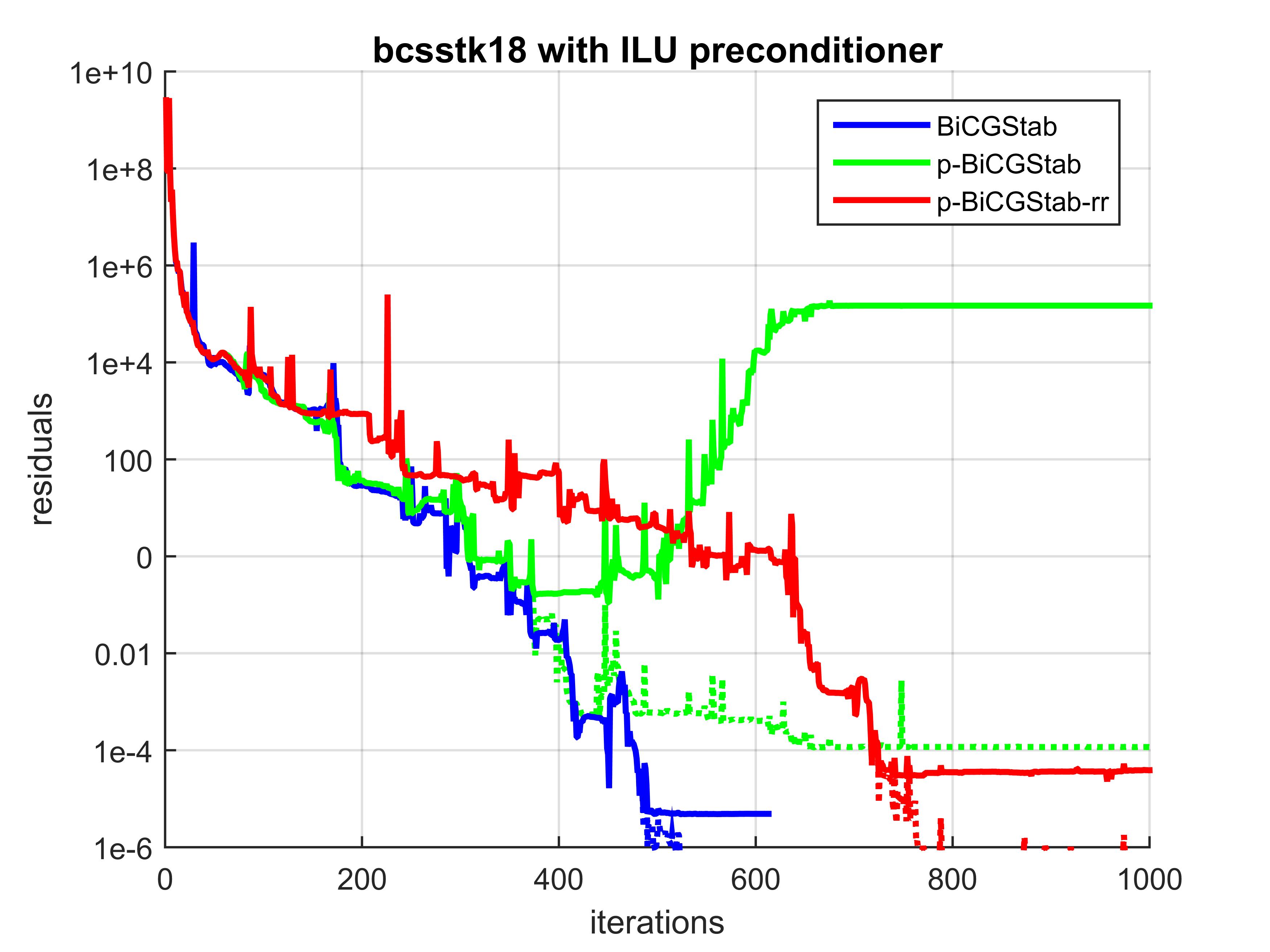} \vspace{0.2cm} \\ 
\includegraphics[width=0.48\textwidth]{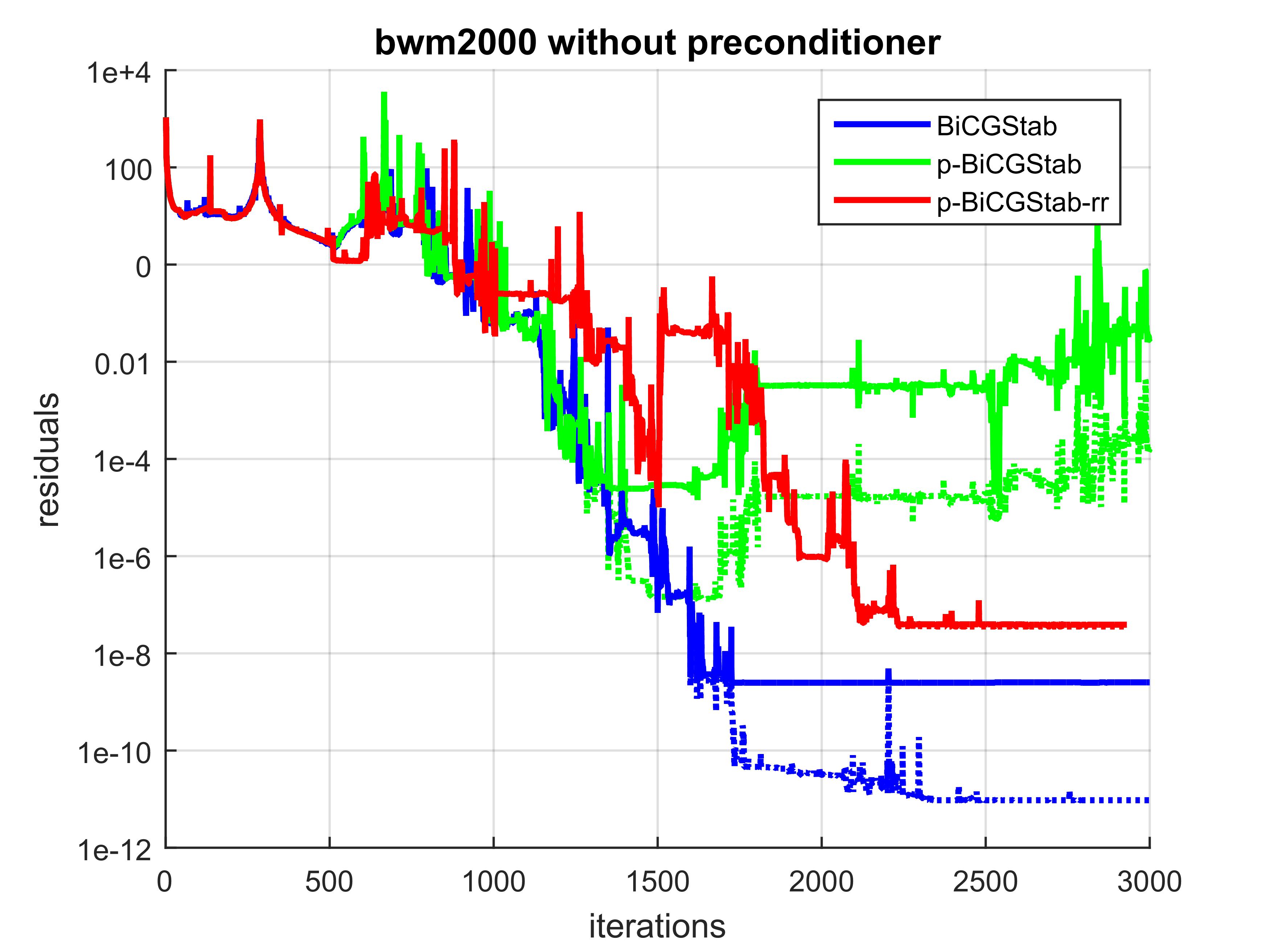} &
\includegraphics[width=0.48\textwidth]{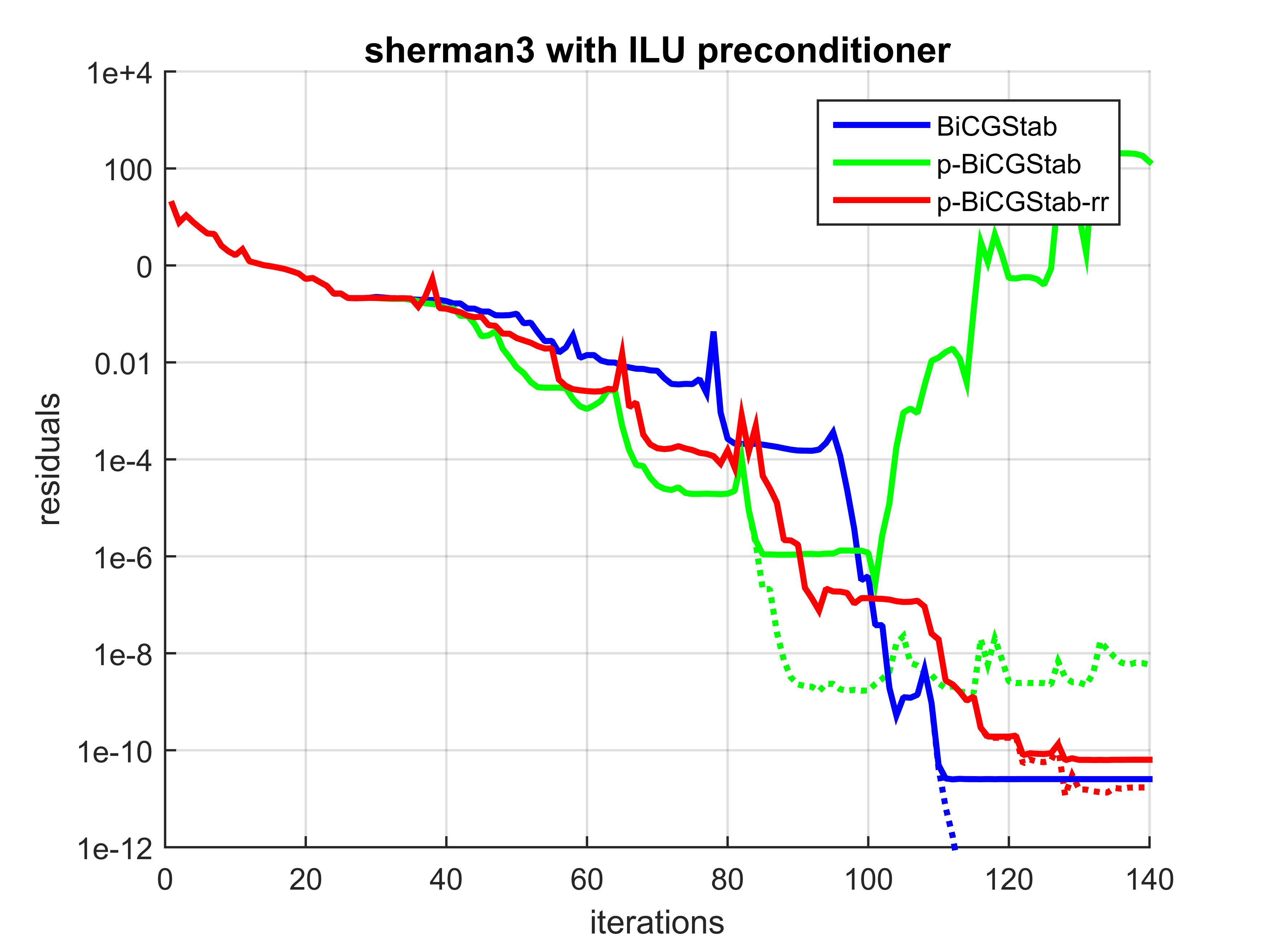}
\end{tabular}
\end{center}
\caption{Residual history of BiCGStab (blue), p-BiCGStab (green) and p-BiCGStab-rr (red) for different test matrices. 
Specifications: see Table \ref{tab:matrix_market2}. 
The dotted lines are the norms of the recursive residuals $\|r_i\|_2$, while solid lines represent the true residual norms $\|b-Ax_i\|_2$. 
The p-BiCGStab residuals stagnate several orders of magnitude above the standard BiCGStab residuals.
The residual replacement strategy improves the attainable accuracy of the p-BiCGStab method.
\label{fig:conv_examples2}}
\end{figure}

Some applications demand a significantly higher accuracy on the solution than imposed by the relatively mild stopping criterion $\|r_i\|_2/\|r_0\|_2 \leq 10^{-6}$ used in Section \ref{sec:matrix}. Table \ref{tab:matrix_market2} shows the maximal attainable accuracy, i.e., the minimal true residual norm $\min_i \|b-Ax_i\|_2$, and the corresponding number of iterations required to attain this accuracy for the BiCGStab and p-BiCGStab algorithms on the Matrix Market selection. The pipelined method is unable to reach the same maximal accuracy as standard BiCGStab. This is a known issue related to most communication-avoiding algorithms, see \cite{carson2014residual}, and also affects pipelined Krylov subspace methods, cf.~\cite{cools2016rounding,ghysels2014hiding}. 

Several orders of magnitude on maximal attainable precision are typically lost when switching to the pipelined algorithm. The accuracy loss is caused by the increased number of \textsc{axpy} operations in the pipelined Alg.~\ref{algo::bicgstab5}, which features $11$ vector recurrence operations, compared to only $4$ recurrences in the standard preconditioned BiCGStab Alg.~\ref{algo::bicgstab4}. Additional recurrence relations are prone to introducing and amplifying local rounding errors in the algorithm, causing the residuals to level off sooner, as has been analyzed in a number of papers among which \cite{cools2016rounding,greenbaum1997estimating,gutknecht2000accuracy,meurant2006lanczos,strakovs2002error,tong2000analysis}.

Although in practice the loss of maximal attainable accuracy is only problematic when a very high precision on the solution is required, from a numerical point of view it is nonetheless important to address this issue. Countermeasures in the form of a residual replacement strategy \cite{carson2014residual,greenbaum1997estimating,sleijpen1996reliable,sleijpen2001differences,van2000residual} are introduced to negate the effect of propagating rounding errors introduced by the additional recurrences in the pipelined method. The residual replacement technique resets the residuals $r_i$ and $\hat{r}_i$, as well as the auxiliary variables $w_i$, $s_i$, $\hat{s}_i$ and $z_i$, to their true values every $k$ iterations, i.e.,
\begin{align*}
	r_i &:=  b-Ax_i, 		  & \hat{r}_i &:= M^{-1}r_i, & w_i &:= A\hat{r}_i, \\
	s_i &:=  A \hat{p}_i, & \hat{s}_i &:= M^{-1}s_i, & z_i &:= A\hat{s}_i. 
\end{align*}
This induces an extra computational cost of 4 \textsc{spmv}s and 2 preconditioner applications every $k$ iterations. However, it is typically sufficient to perform only a small number of residual replacement steps to improve the maximal attainable accuracy, as indicated by the average \#nrr percentage displayed in the bottom right corner of Table \ref{tab:matrix_market2}. Hence, the replacement period $k$ is often large, limiting the computational overhead of performing the residual replacements. In Section \ref{sec:parallel} it is shown that the extra \textsc{spmv} cost is negligible compared to the global cost of the method. The pipelined BiCGStab method with residual replacement is denoted as p-BiCGStab-rr for short.

By resetting the residual to its true value every $k$-th iteration, any build-up rounding errors are effectively eliminated. Through the periodic removal of rounding errors the algorithm is able to attain the same maximal accuracy as standard BiCGStab. This is illustrated by the p-BiCGStab-rr residual norms in Table \ref{tab:matrix_market2}. However, the introduction of the residual replacement step alters the p-BiCGStab convergence behavior, resulting in an increase in iterations required to attain the maximum precision on the solution for some problems. This delayed convergence \cite{strakovs2002error} is clearly visible in Figure \ref{fig:conv_examples2}, where the convergence histories (true residual norms) of the BiCGStab and p-BiCGStab methods on four selected matrices are shown up to a high accuracy. 

Table \ref{tab:matrix_market2} indicates an average reduction of iterations by $11.0\%$ for p-BiCGStab compared to standard BiCGStab over all listed matrices. 
These iteration numbers are however related to the lower maximal attainable accuracy of p-BiCGStab, and in general do not imply p-BiCGStab converges faster than the original BiCGStab method, cf.~Fig.~\ref{fig:conv_examples2}.
For the p-BiCGStab-rr method an average increase in iterations of $22.1\%$ is observed. Indeed, for some problems such as \texttt{1138\_bus}, \texttt{fs\_760\_3} and \texttt{utm5940} a significant increase in iterations and convergence irregularity is observed, see also Fig.~\ref{fig:conv_examples2}. Final residuals displayed in Table \ref{tab:matrix_market2} show that a high accuracy on the solution \emph{can} be achieved by use of the residual replacement strategy when required by the application. The associated increase in iterations implies a trade-off between speedup and accuracy that should be kept in mind when using the p-BiCGStab-rr method.

Note that the replacement period parameter is chosen manually for the matrices in Table \ref{tab:matrix_market2}, based on an ad hoc estimation of the total number of BiCGStab iterations. These values are relatively arbitrary, and a different choice of this parameter could lead to either significantly slower or faster convergence of the p-BiCGStab-rr algorithm.

Figure \ref{fig:conv_examples2} illustrates an additional beneficial effect of the residual replacement strategy on the pipelined BiCGStab method. The convergence history of the p-BiCGStab method is generally comparable (albeit not identical) to the BiCGStab residuals, up to the stagnation point where p-BiCGStab attains maximal accuracy. However, it is observed that after some iterations the p-BiCGStab true residuals start to increase again. 
Standard BiCGStab does not display this unwanted behavior. By periodically resetting the residual and auxiliary variables to their true values, the residual replacement strategy resolves these robustness issues. Figure \ref{fig:conv_examples2} shows a stagnation of the p-BiCGStab-rr residual norm similar to standard BiCGStab after maximal accuracy is attained. 

\begin{figure}[t]
\begin{center}
\begin{tabular}{cc}
\includegraphics[width=0.40\textwidth]{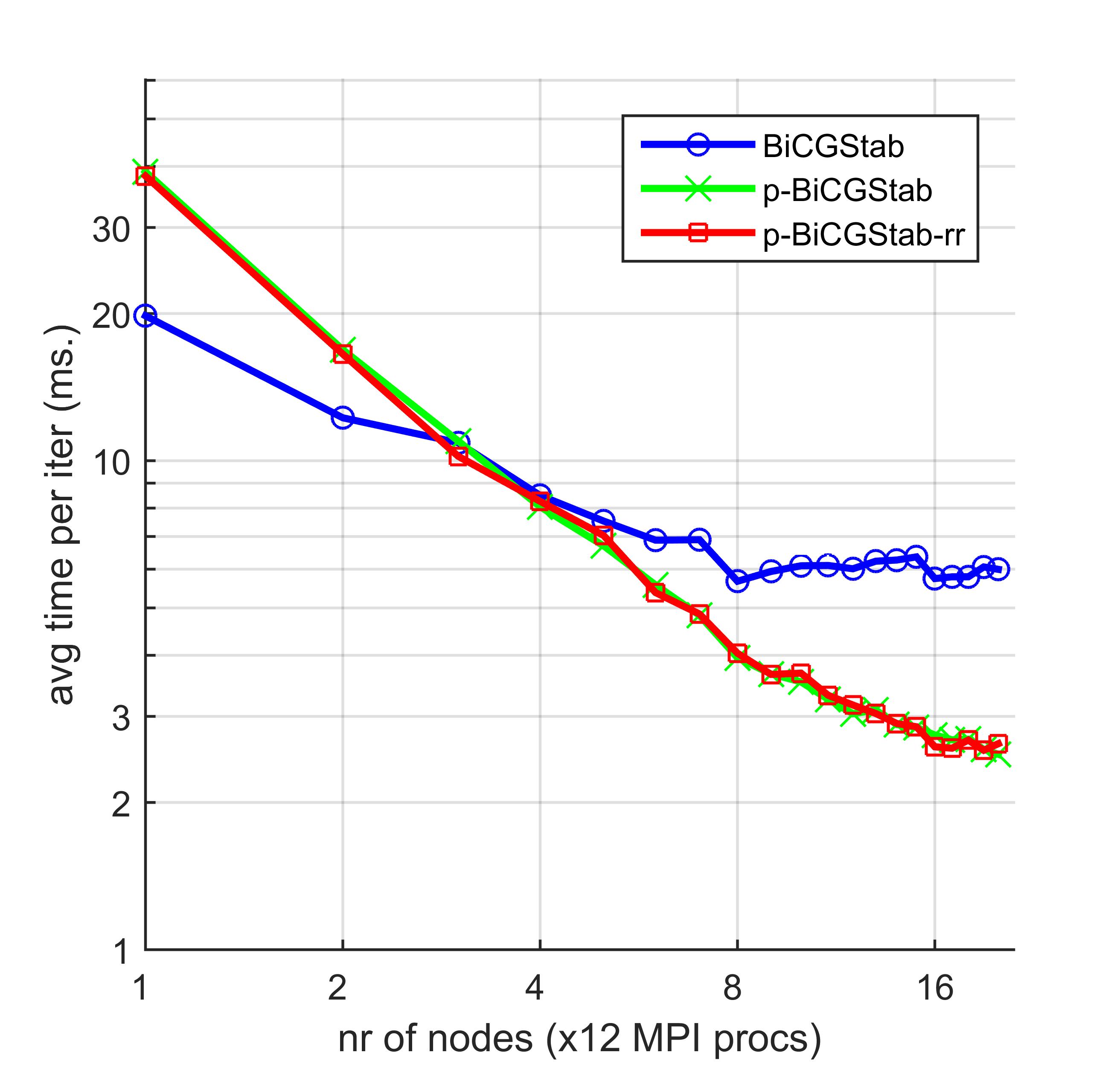} &
\includegraphics[width=0.40\textwidth]{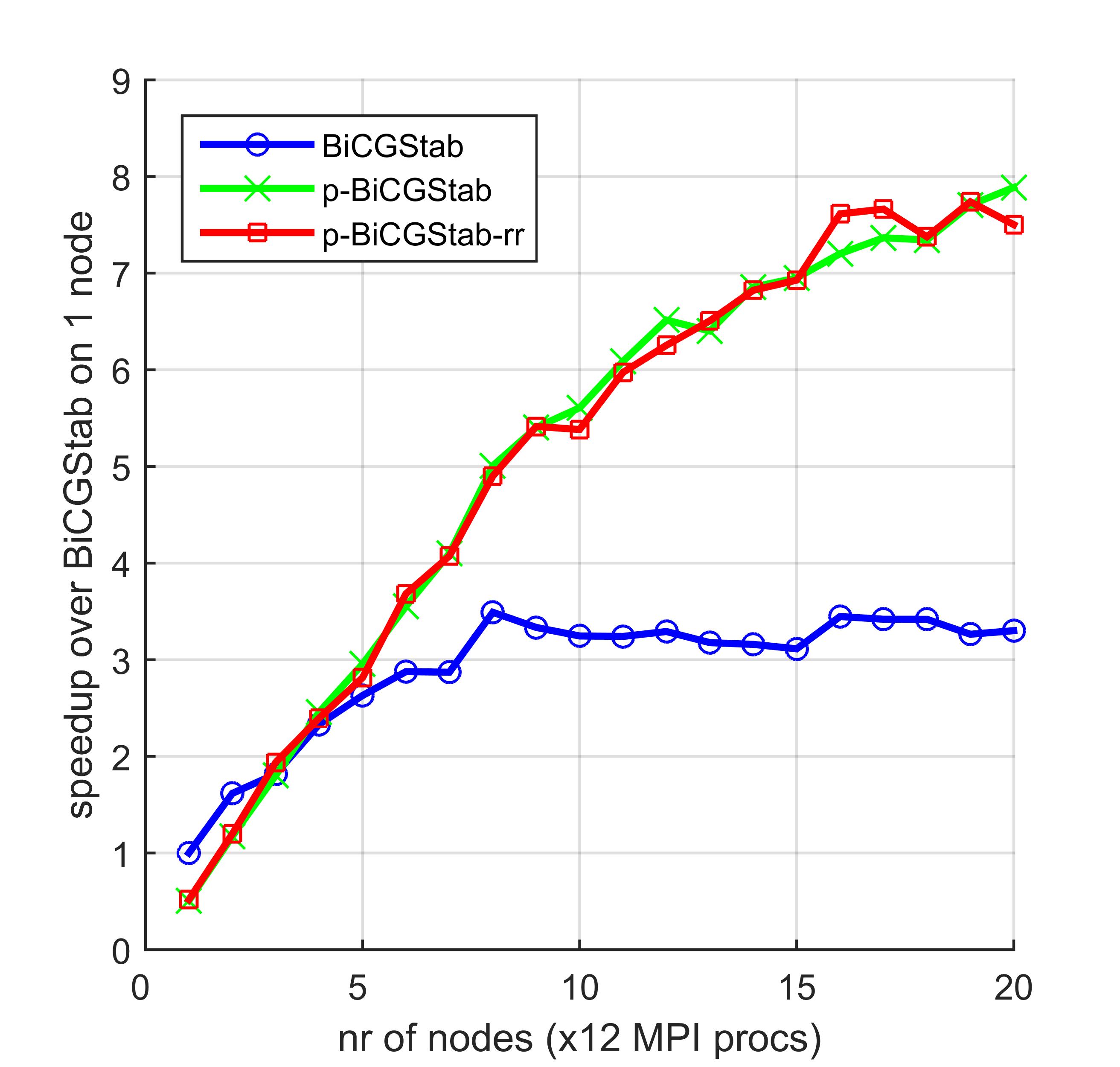} \\
\includegraphics[width=0.40\textwidth]{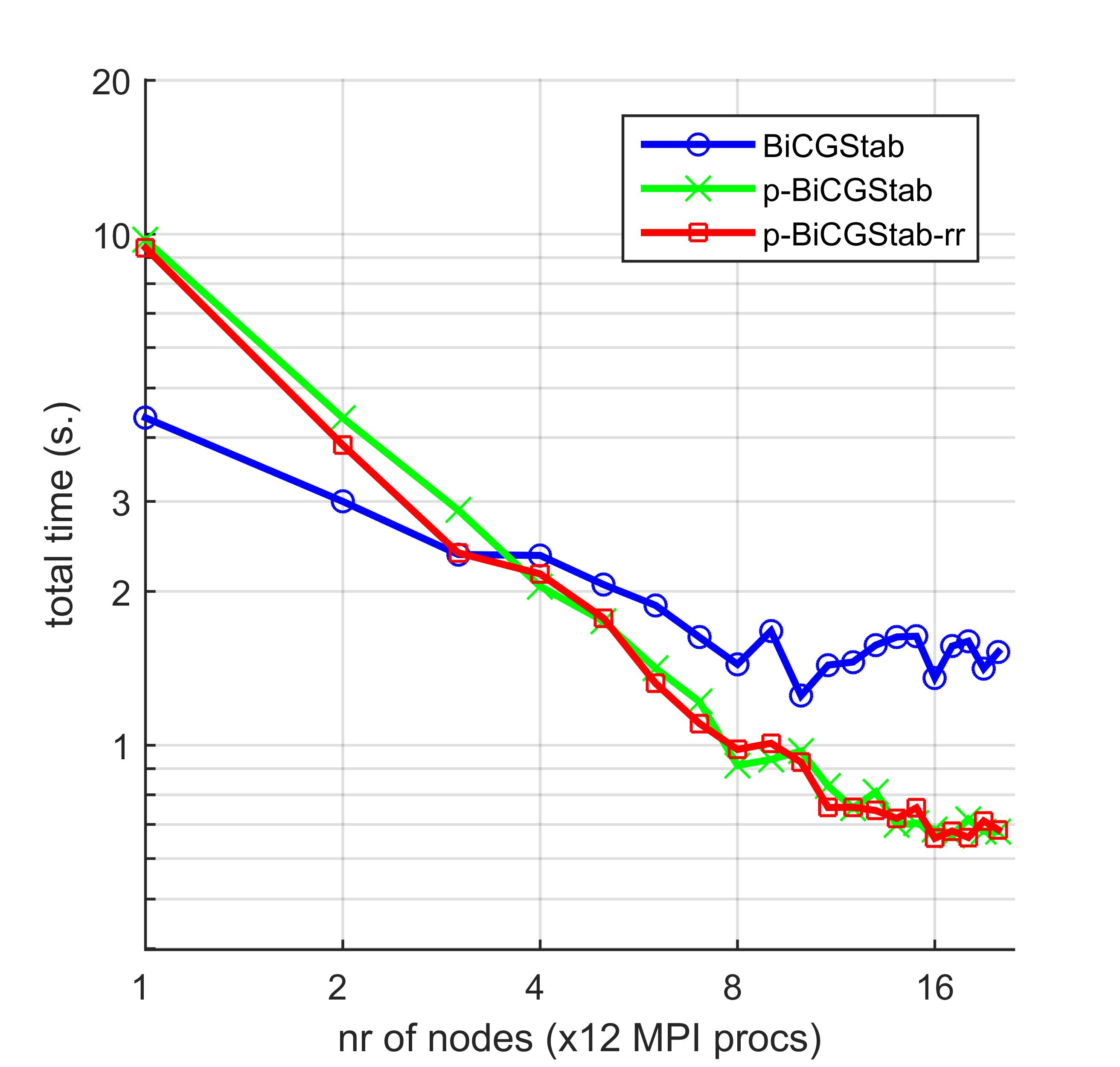} &
\includegraphics[width=0.40\textwidth]{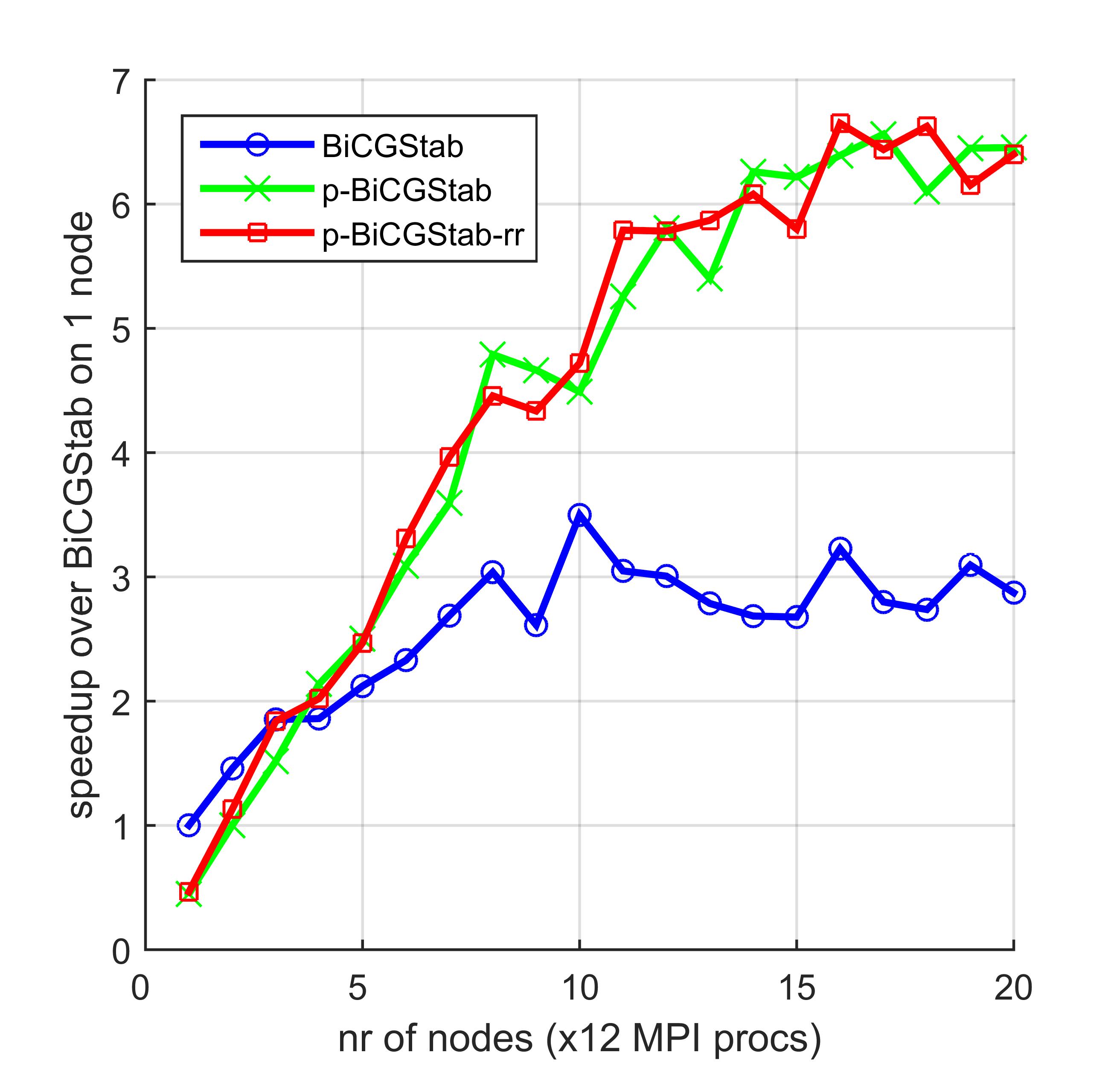} 
\end{tabular}
\end{center}
\caption{\textbf{(PTP1)} Strong scaling experiment on up to 20 nodes (240 cores). 
Top left: Average time per iteration (\texttt{log10} scale) as function of the number of nodes (\texttt{log2} scale). 
Top right: Speedup (per iteration) over standard BiCGStab on a single node. 
Bottom left: Total CPU time as function of the number of nodes. 
Bottom right: Absolute speedup over standard BiCGStab on a single node.
All methods were set to converge to a tolerance of $10^{-6}$ on the scaled residual, which was reached in 205 (min.) to 282 (max.) iterations. The p-BiCGStab-rr algorithm performs a replacement step every 100 iterations. 
\label{fig:timings}}
\end{figure}

\section{Parallel performance} \label{sec:parallel}

\begin{figure}[t]
\begin{center}
\begin{tabular}{cc}
\includegraphics[width=0.40\textwidth]{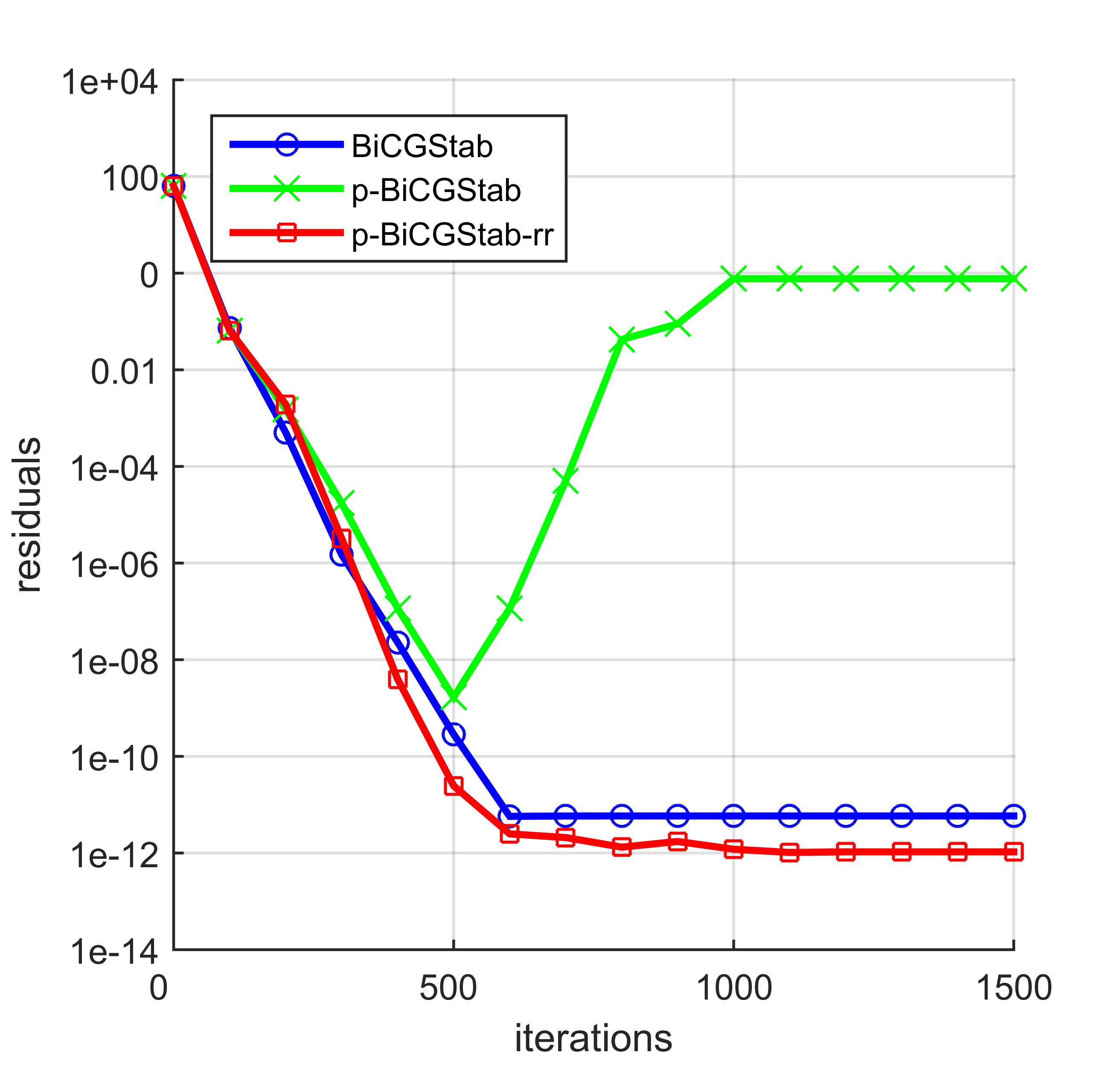} &
\includegraphics[width=0.40\textwidth]{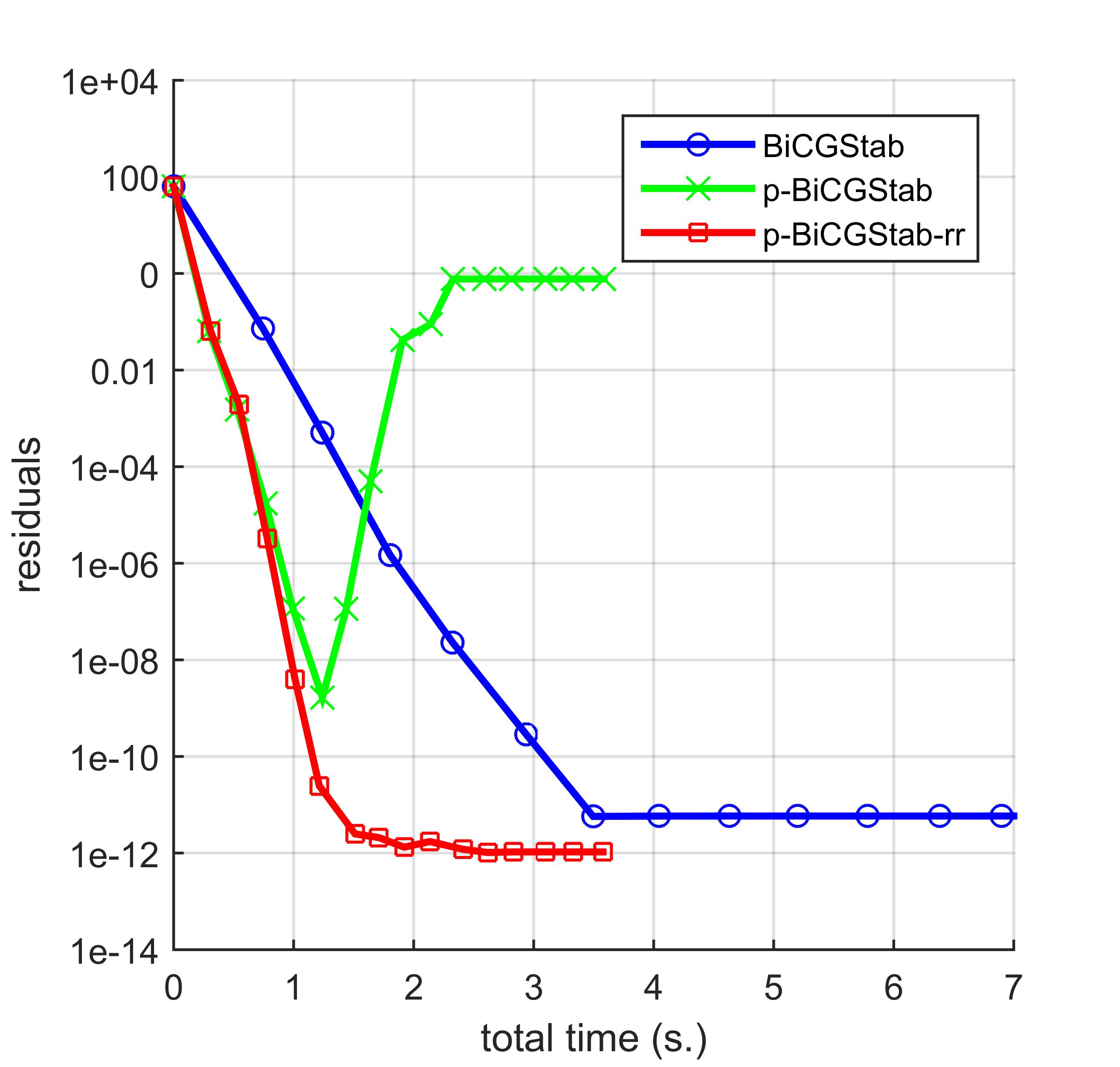} 
\end{tabular}
\end{center}
\caption{\textbf{(PTP1)} Accuracy experiment on 20 nodes (240 cores). 
Left: True residual norm $\|b-Ax_i\|_2$ as function of iterations. 
Right: True residual norm as function of total time spent by the algorithm. 
Maximal number of iterations is 2000 for all methods. The p-BiCGStab-rr algorithm performs a replacement step every 100 iterations (max.~10 replacements).
\label{fig:timings2}}
\end{figure}

In this section we illustrate the parallel performance of the 
pipelined BiCGStab method, Alg.~\ref{algo::bicgstab5}. 
The scaling and accuracy experiments in this section are performed on a small cluster with 20 compute nodes, consisting of two 6-core Intel Xeon 
X5660 Nehalem 2.80 GHz processors each (12 cores per node), for a total of 240 cores. Nodes are connected by $4\,\times\,$QDR 
InfiniBand technology, providing 32 Gb/s of point-to-point bandwidth for message passing and I/O.
Since each node consists of 12 cores, we use 12 MPI processes per node to fully exploit parallelism on the machine.
The MPI library used for this experiment is MPICH-3.1.3\footnote{\url{http://www.mpich.org/}}. The environment variables 
\texttt{MPICH\_ASYNC\_PROGRESS=1} and \texttt{MPICH\_MAX\_THREAD\_SAFETY=multiple} are set to ensure optimal 
parallelism\footnote{We point out that the need for these settings depends on the specific hard- and firmware used in practice.};
the first variable enables asynchronous non-blocking reductions, while the second allows for a process to have multiple threads
which simultaneously call MPI functions.

\textbf{Parallel test problem 1.} The pipelined BiCGStab Alg.~\ref{algo::bicgstab5} was implemented in the open-source PETSc library \cite{petsc-web-page}, version 3.6.2, as a direct modification of the \texttt{fbcgs} implementation found in the PETSc Krylov solvers (KSP) folder. The first benchmark problem 
is a 2D PDE-type model, defined by the unsymmetric 5-point stencil
\begin{equation}A_1^{st} =
\begin{bmatrix}
     & -1  &  \\
    -1 & ~~4 & -\varepsilon \\
      & -\varepsilon  &  \\
 \end{bmatrix}, \quad \varepsilon = 1 - 0.001.
\tag{PTP1}
\end{equation}
The right-hand side $b = A_1\hat{x}$ is constructed using the exact solution $\hat{x} = \bold{1}$. The operator is a modified 2D Poisson PDE stencil, discretized using second order finite difference approximations on a uniform grid, which is available in the PETSc distribution as example $2$ in the Krylov solvers folder. Although academic, the model problem (PTP1) is non-trivial from an iterative solver perspective, since a large number of the operator's eigenvalues are located close to zero. The number of grid points is set to 1.000 per spatial dimension, resulting in a total of one million unknowns. 

As illustrated in Section \ref{sec:precpipebicgstab}, the inclusion of a preconditioner in pipelined algorithms is generally straightforward. However, to efficiently overlap the preconditioner application with the global communication phase, the preconditioner itself should not be bottle-necked by communication. As such, a simple block Jacobi preconditioner is trivially well-suited for this purpose, whereas the inclusion of a more advanced preconditioning scheme like parallel ILU \cite{chow2015fine} or additive Schwarz \cite{smith2004domain} requires a more careful treatment. For simplicity no preconditioner is included in the following experiment.  

Figure \ref{fig:timings} (top left) shows the average time per iteration required to solve the problem up to a tolerance of $10^{-6}$ on the scaled residual as a function of the number of nodes. For this test problem and hardware configuration, pipelined BiCGStab method (green) starts to outperform standard BiCGStab (blue) when the number of nodes exceeds four. Figure \ref{fig:timings} (top right) shows the same data, reformulated as speedup over standard 1-node BiCGStab. The p-BiCGStab method scales well up to 20 nodes. The bottom row in Figure \ref{fig:timings} shows the non-averaged total time (bottom left) and absolute speedup in function of the number of nodes (bottom right), which display similar scaling behavior, albeit slightly less smooth due to the varying number of iterations between individual runs.
The maximum speedup on 20 nodes over standard BiCGStab on a single node is $7.89\times$ (computed on averaged timings). In contrast, for the model problem and hardware configuration in this benchmark experiment, the standard BiCGStab method stops scaling at around 8 nodes, obtaining a speedup of only $3.30\times$ on 20 nodes. Hence, pipelined BiCGStab attains a net speedup (per iteration) of $2.39\times$ compared to standard CG when both are executed on 20 nodes, which approximates the theoretically optimal speedup of $2.5\times$, cf.~Section \ref{sec:ibicgstab}. The absolute (non-averaged) speedup factor of p-BiCGStab (270 iterations) over standard BiCGStab (254 iterations) on 20 nodes is $2.25\times$. Performance results for the p-BiCGStab-rr method are similar to those of p-BiCGStab. The small computational overhead from the residual replacement steps does not affect strong scaling.

Figure \ref{fig:timings2} shows the accuracy of the solution in function of the number of iterations (left) and in function of the total computational time (right) for the BiCGStab and p-BiCGStab algorithms on the 2D unsymmetric benchmark problem on a 20 node setup. Standard BiCGStab attains a maximal accuracy on the solution corresponding to a residual 2-norm of 5.8e-12 in 4.0 seconds. The pipelined variant is significantly faster, attaining a residual norm of 1.7e-9 in only 1.2 seconds. However, for the pipelined method a higher accuracy is only obtainable by including the residual replacement strategy. The p-BiCGStab-rr method is able to attain a residual norm of 2.5e-12 in 1.5 seconds, which is significantly faster than standard BiCGStab for a comparable accuracy.

\begin{figure}[t]
\begin{center}
\begin{tabular}{cc}
\includegraphics[width=0.40\textwidth]{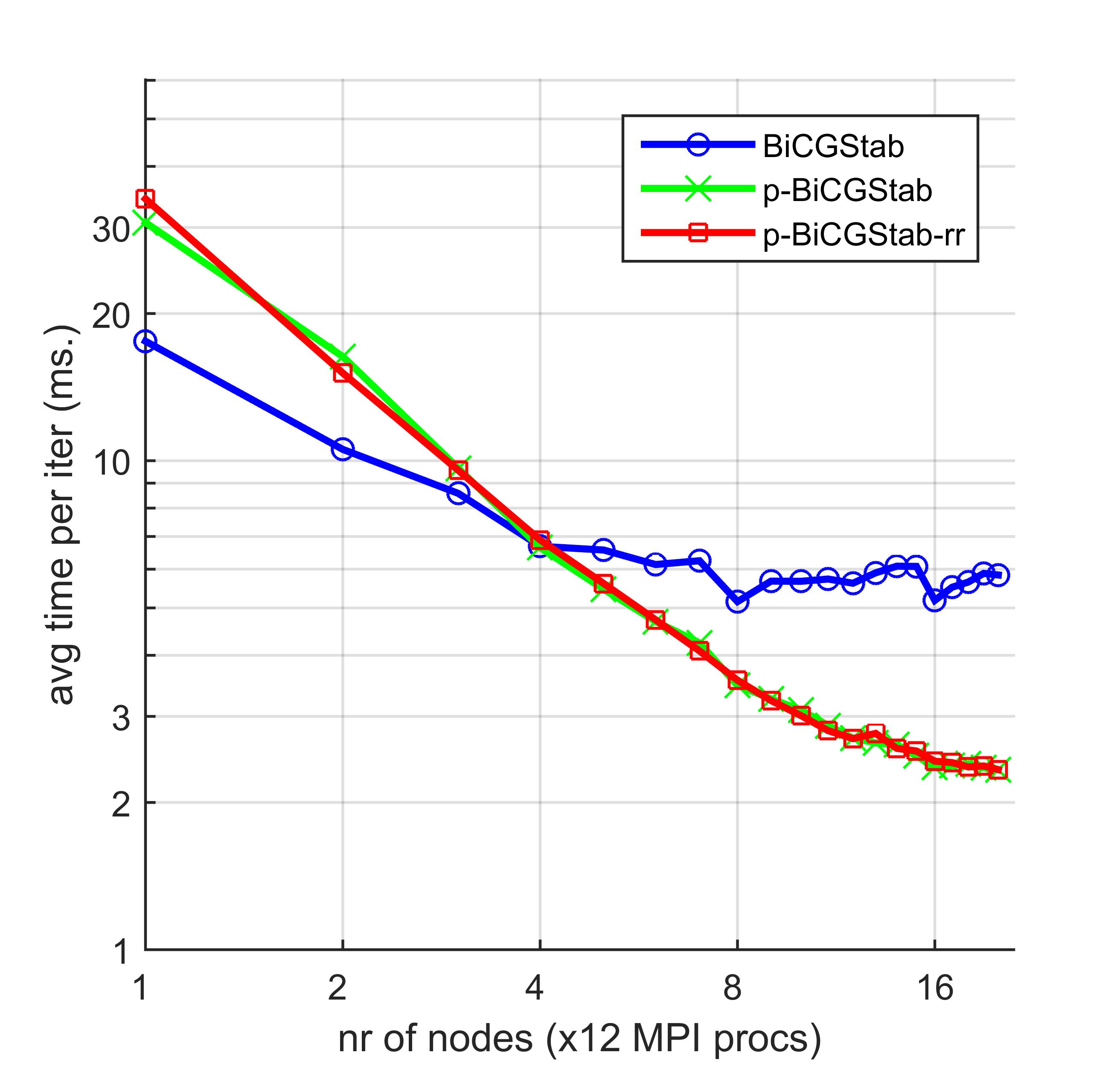} &
\includegraphics[width=0.40\textwidth]{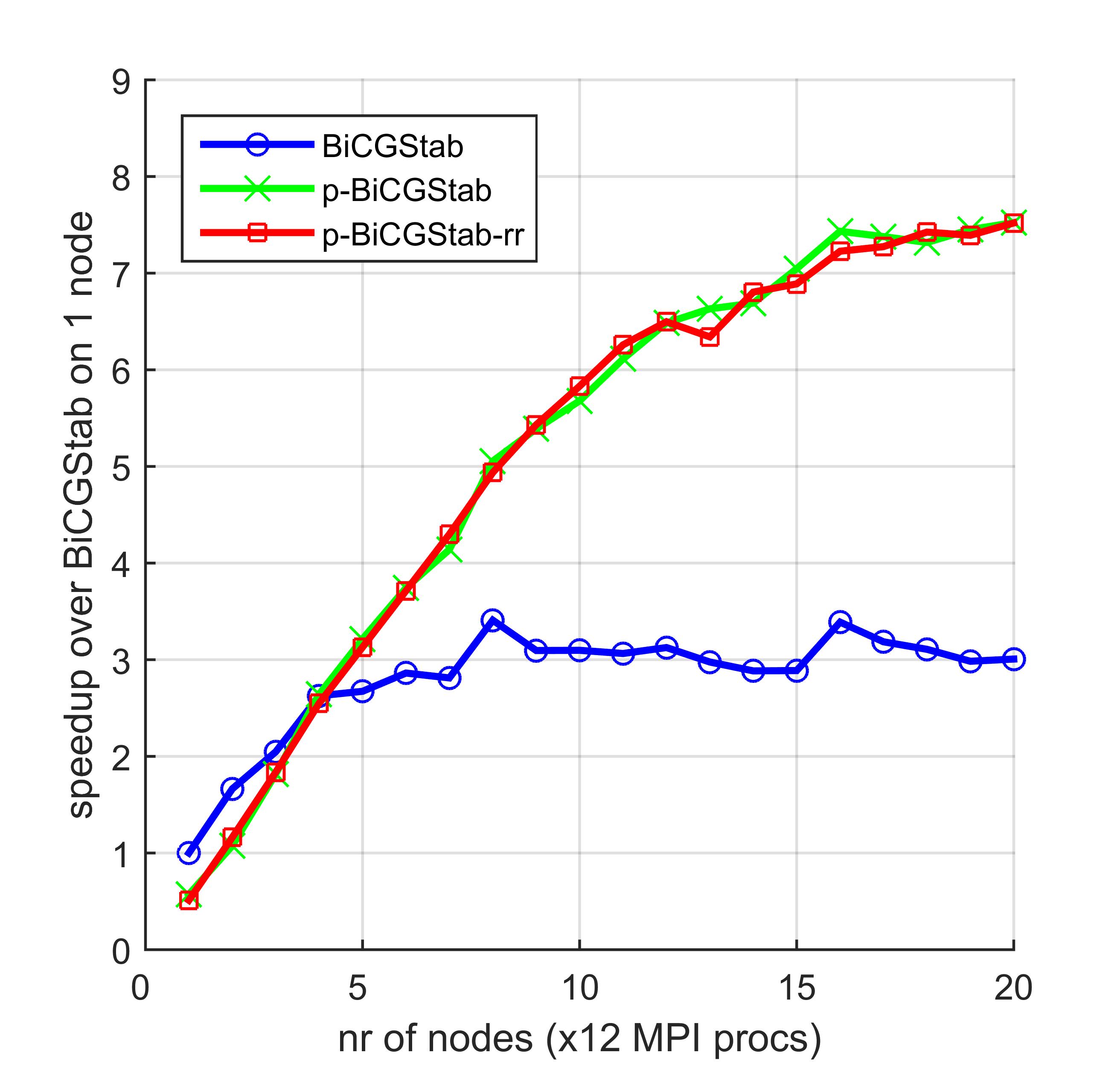} \\
\includegraphics[width=0.40\textwidth]{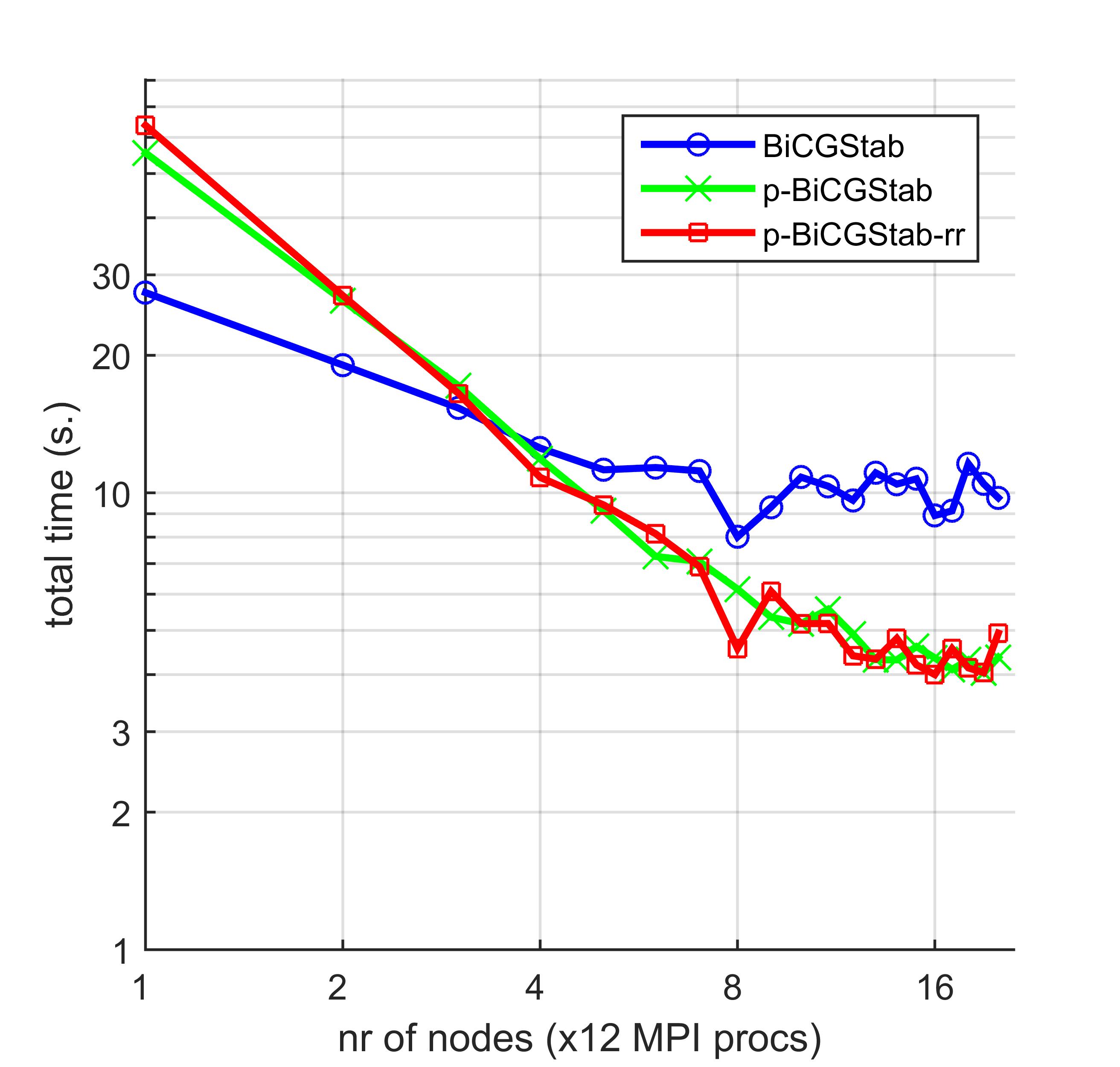} &
\includegraphics[width=0.40\textwidth]{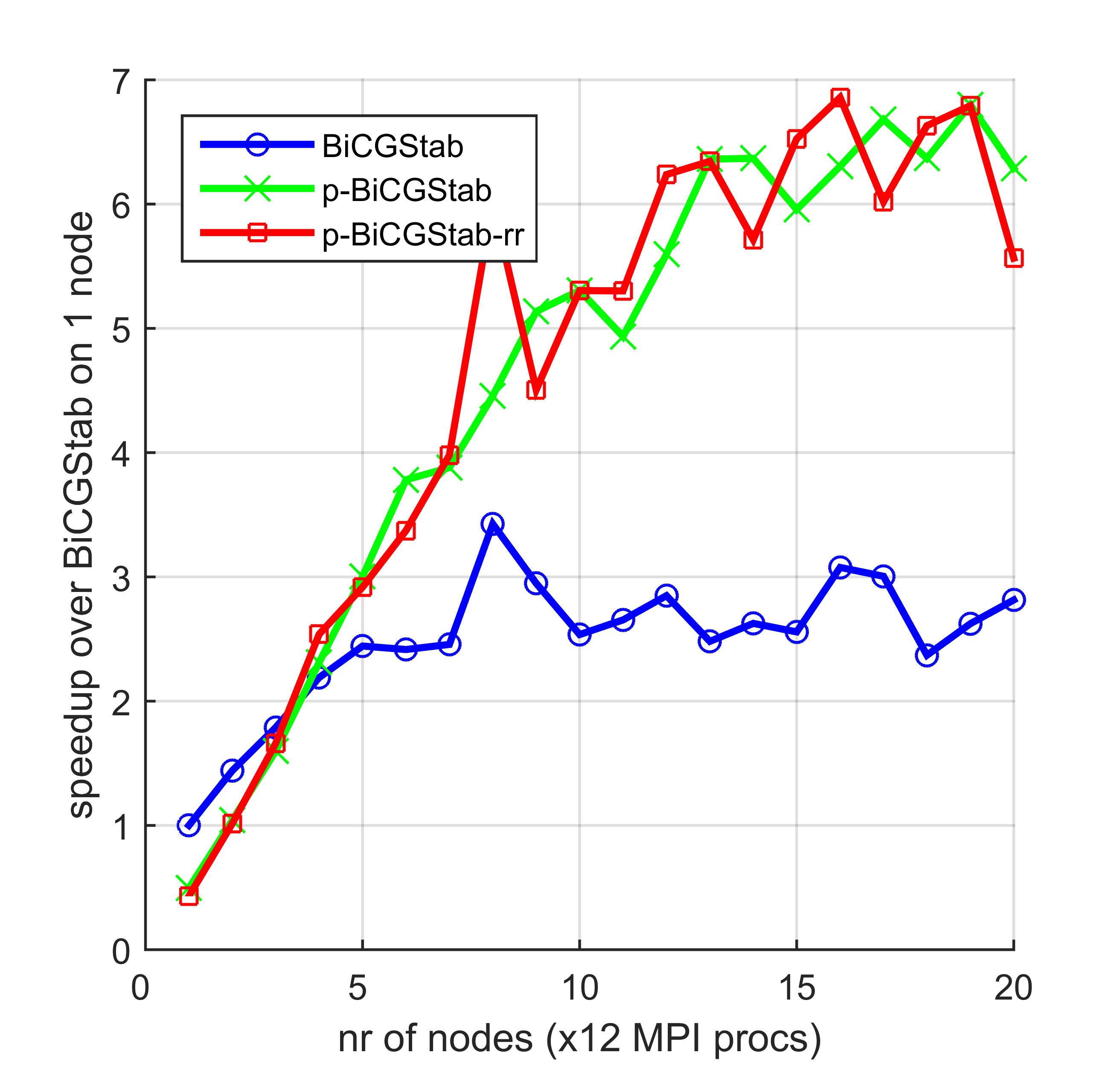} 
\end{tabular}
\end{center}
\caption{\textbf{(PTP2)} Strong scaling experiment on up to 20 nodes (240 cores). 
Top left: Average time per iteration (\texttt{log10} scale) as function of the number of nodes (\texttt{log2} scale). 
Top right: Speedup (per iteration) over standard BiCGStab on a single node. 
Bottom left: Total CPU time as function of the number of nodes. 
Bottom right: Absolute speedup over standard BiCGStab on a single node.
All methods converged to a scaled residual tolerance of $10^{-6}$, which was reached in 1283 (min.) to 2112 (max.) iterations. The p-BiCGStab-rr algorithm performs a replacement step every 100 iterations (max.~10 replacements). 
\label{fig:timings3}}
\end{figure}

\textbf{Parallel test problem 2.} The second benchmark problem used to asses parallel performance is a Helmholtz-type PDE model given by the 5-point stencil
\begin{equation}A_2^{st} =
\begin{bmatrix}
     & -1  &  \\
    -1 & ~~1 & -1 \\
      & -1  &  \\
 \end{bmatrix},
\tag{PTP2}
\end{equation}
and a right-hand side $b = A_2\hat{x}$, where again $\hat{x} = \bold{1}$. It can be considered as a 2D Poisson operator, shifted by a diagonal matrix that relates to the Helmholtz wave number. This operator is highly indefinite and notably hard to solve using iterative methods, see \cite{ernst2012difficult}. The one million unknowns system $A_2 x = b$ is solved by unpreconditioned\footnote{Most standard preconditioners based on e.g.~incomplete LU factorization, multigrid methods or domain decomposition methods available in PETSc do not improve convergence for Parallel test problem 2. The authors are aware of the existence of specialized preconditioning techniques for Helmholtz-type problems, as proposed in e.g.~\cite{engquist2011sweeping,erlangga2004class}. However, a discussion on the effectiveness of these preconditioners is beyond the scope of this work.} BiCGStab, Alg.~\ref{algo::bicgstab1}, and its pipelined variant, Alg.~\ref{algo::bicgstab3}.

Figure \ref{fig:timings3} shows timing (left) and speedup (right) results for test problem (PTP2) on one up to 20 nodes. The averaged speedup graph (top right) for p-BiCGStab is largely comparable to the results for (PTP1), showing good scaling on up to 20 nodes with a per-iteration speedup of $7.52\times$ (non-averaged: $6.29\times$) compared to 1-node BiCGStab. 
Total timings are slightly more oscillating due to the significant differences in iterations between individual runs, see Table \ref{tab:iters}. 
The total time spent by the p-BiCGStab algorithm on 20 nodes is 9.8 s.~(for 1670 iterations), whereas the p-BiCGStab algorithm requires only 4.6 s.~(for 1870 iterations) to attain the same accuracy. Hence, pipelining results in a net speedup factor of $2.23\times$ on 20 nodes for this model problem.

\begin{table}[t]
\centering
\vspace{1.0cm}
\scriptsize
\begin{tabular}{| l | c c c c c c c c c c |}
\hline 
 nodes & 1 & 2 & 3 & 4 & 5 & 6 & 7 & 8 & 9 & 10 \\ 
\hline 
 BiCGStab      & 1563 & 1805 & 1789 & 1875 & 1710 & 1852 & 1789 & 1555 & 1641 & 1909 \\
 p-BiCGStab    & 1807 & 1614 & 1779 & 1787 & 1673 & 1547 & 1668 & 1773 & 1640 & 1673 \\
 p-BiCGStab-rr & 1857 & 1788 & 1728 & 1570 & 1677 & 1721 & 1688 & 1283 & 1884 & 1718 \\
\hline
 nodes & 11 & 12 & 13 & 14 & 15 & 16 & 17 & 18 & 19 & 20 \\
\hline 
 BiCGStab      & 1805 & 1715 & 1875 & 1717 & 1765 & 1722 & 1657 & 2050 & 1778 & 1670 \\
 p-BiCGStab    & 1936 & 1811 & 1629 & 1642 & 1849 & 1843 & 1726 & 1796 & 1713 & 1870 \\
 p-BiCGStab-rr & 1845 & 1628 & 1562 & 1861 & 1650 & 1647 & 1889 & 1750 & 1701 & 2112 \\
\hline
\end{tabular}
\caption{\textbf{(PTP2)} Reference table showing the iterations required to attain the scaled residual tolerance 1e-6, corresponding to a true residual norm $\|b - Ax_i\|_2 \leq$ 3.0e-3, as a function of the number of nodes.}
\label{tab:iters}
\end{table}

\section{Conclusions} \label{sec:conclusions}

Pipelined algorithms (partially) circumvent the traditional global synchronization bottleneck in traditional Krylov subspace methods. As a consequence, they offer better scalability in the strong scaling limit for computing solutions to large and sparse linear systems on massively parallel hardware. In this work we proposed a general framework for the derivation of a pipelined variant of a given Krylov subspace method. 
The pipelining framework consists of two main phases. In the first step, denoted as \emph{communication-avoiding}, the standard Krylov algorithm is rewritten into a mathematically equivalent algorithm with fewer global synchronization points. This is achieved by combining the global reduction phases of different dot-products scattered across the algorithm into one global communication phase. The second step, called \emph{communication-hiding}, subsequently reformulates the algorithm such that the remaining global reduction phases are overlapped by the sparse matrix-vector product and preconditioner application. As such, the typical communication bottleneck is mitigated by hiding communication time behind useful computational work. 

Applications of the proposed framework include the reformulation of several widely used Krylov subspace methods, such as the Conjugate Gradient method (CG) for symmetric and positive definite linear systems \cite{ghysels2014hiding}, and the Generalized Minimal Residual method (GMRES) \cite{ghysels2013hiding} and Bi-Conjugate Gradient Stabilized method (BiCGStab) for the solution of general unsymmetric and/or indefinite systems.
The proposed high-level framework can be used to derive a length-one pipelined version of any Krylov subspace method. The development of a general framework for the derivation of length-$l$ pipelined methods is left as future work.

To illustrate the methodology, the pipelining framework is successfully applied to the BiCGStab method for the solution of large and sparse unsymmetric linear systems. The pipelined BiCGStab method (p-BiCGStab) reduces the number of global synchronization points from three to two, and overlaps the remaining global reduction phases with computational work. This induces a theoretical speed-up of up to 250\% over traditional BiCGStab on a large number of nodes. Contrary to the so-called $s$-step methods \cite{carson2015communication,carson2014residual,carson2013avoiding,chronopoulos1991s,chronopoulos1989s,chronopoulos2010block,chronopoulos1996parallel}, the combination of pipelined methods and preconditioning is straightforward, as is illustrated by the derivation of the preconditioned pipelined BiCGStab method in this work. However, to ensure optimal scaling the chosen preconditioner preferably requires only a limited amount of global communication. 

Numerical experiments on a moderately sized cluster show that the p-BiCGStab method displays significantly increased parallel performance and improved strong scaling compared to standard BiCGStab on an increasing number of computational nodes. In practice a speedup of $2.0$ to $2.5\times$ over the traditional BiCGStab method can be expected when both are executed on the same number of parallel processors.

Finally, the experimental results point out two minor numerical drawbacks that originate from reordering the BiCGStab algorithm into a pipelined version. In the extremely small residual regime, a loss of maximal attainable accuracy can be expected, which is a typical phenomenon related to pipelined (and other communication-avoiding) Krylov subspace methods \cite{cools2016rounding,ghysels2014hiding}. 
Furthermore, due to the introduction of additional \textsc{axpy} operations by the pipelining framework, the p-BiCGStab algorithm is typically less robust with respect to numerical rounding errors compared to the standard algorithm. It is observed that the p-BiCGStab residuals are not guaranteed to remain at the same level after attaining the maximal accuracy, which is a highly unwanted feature. 
Both of these numerical issues are simultaneously resolved by including a residual replacement strategy \cite{carson2014residual,greenbaum1997estimating,sleijpen1996reliable,sleijpen2001differences,van2000residual} in the pipelined method. Indeed, through a periodical reset of the residual and auxiliary variables to their true values by explicitly computing the corresponding \textsc{spmv}s, it is shown that both robustness and attainable accuracy can be restored to the original BiCGStab method's level at the expense of a moderate added computational cost.

\section{Acknowledgments} This work is partially funded by the EXA2CT European Project on
Exascale Algorithms and Advanced Computational Techniques, which receives 
funding from the EU's Seventh Framework Programme (FP7/2007-2013) under grant agreement no.~610741.
Additionally, S.\,C.\,is funded by the Research Foundation Flanders (FWO). 
The authors are grateful to Bram Reps for his preliminary work on this topic, and to Pieter Ghysels and Patrick Sanan for their useful insights and remarks.

{\footnotesize
\bibliographystyle{plain}
\bibliography{refs}
}

\end{document}